\documentclass[12pt]{article}
\usepackage{amsmath,amsthm, amssymb}
\usepackage{graphicx}
\usepackage{leqno}
\usepackage{hyperref}
\numberwithin{equation}{section}
\begin{document}
	\title{\vskip-40pt Wave Function Collapse and the No-Superluminal-Signaling Principle}
	\author{Edward J. Gillis\footnote{email: gillise@provide.net}}

\providecommand{\keywords}[1]{\textbf{Keywords} #1}
	
	\maketitle
	
	\begin{abstract} 
		
\noindent 
The assumption that wave function collapse is a real occurrence has very interesting consequences - both experimental and theoretical. Besides predicting observable deviations from linear evolution, it implies that these deviations must originate in nondeterministic effects at the elementary level in order to prevent superluminal signaling, as demonstrated by Gisin. This lack of determinism implies that information cannot be instantiated in a reproducible form in isolated microsystems (as illustrated by the No-cloning theorem). By stipulating that information is a reproducible and referential property of physical systems, one can formulate the no-signaling principle in strictly physical terms as a prohibition of the acquisition of information about spacelike-separated occurrences. This formulation provides a new perspective on the relationship between relativity and spacetime structure, and it imposes tight constraints on the way in which collapse effects are induced. These constraints indicate that wave function collapse results from (presumably small) nondeterministic deviations from linear evolution associated with nonlocally entangling interactions. This hypothesis can be formalized in a stochastic collapse equation and used to assess the feasibility of testing for collapse effects. 
\newline

\noindent\keywords{quantum measurement, wave function collapse, no-signaling, relativity, nondeterminism, information}

\end{abstract}

\section{Introduction}
\label{intro}

Is wave function collapse a \textit{physical} occurrence? Discussions about foundational issues often ignore the fact that this is an empirical question. The clear inconsistency between the experimental predictions of projection and those of linear evolution was emphasized by Bell\cite{Bell_CH}. Using the Coleman-Hepp model\cite{Hepp} he constructed complex observables to illustrate the difference between the effects of collapse and the persistence of superposition: \begin{quotation}\noindent
	``So long as nothing, in principle, forbids the consideration of 
	such arbitrarily complicated observables, it is not permitted to speak 
	of wave packet reduction. While for any given observable one can find 
	a time for which the unwanted interference is as small as you like, for 
	any given time one can find an observable for which it is as big as you 
	do \textit{not} like.''\end{quotation}

The fact that the projection postulate is testable in principle has important implications for debates about quantum foundations. First, it immediately undercuts the claim that it makes no difference where one draws the line between the measured system and the measuring apparatus. This claim is central to the Copenhagen 
interpretation\cite{Bohr_EPR,Heisenberg_a,Heisenberg_b}. Second, it shows that various interpretations \textit{can be falsified} depending on whether or not they predict (or permit) wave function collapse. No-collapse interpretations such as that of Everett\cite{Everett}, decoherence accounts\cite{Zurek_Decoh,Hartle}, 
or pilot wave theories\cite{deBroglie,Bohm,Bohm_Hiley} would be falsified by clear evidence of the breakdown of superposition, and collapse proposals such as that of Ghirardi, Rimini, and Weber (GRW)\cite{GRW} and related hypotheses\cite{Ghirardi_Bassi,Pearle_1,Pearle_2,Bedingham,Gao_1} would be undermined if perfect superposition could be shown to persist to arbitrarily large scales of interaction.

The inconsistency (or ambiguity) in the predictions concerning measurement outcomes means there is a fundamental \textit{logical} problem at the core of contemporary physics. The difficulty in carrying out experiments that could actually test the predictions should not be used as an excuse to ignore the logical gaps in current 
theory. The experiments might be difficult, but demonstrating the inconsistencies is very easy. To highlight this point Section 2 describes a modified version of Bell's demonstration in \cite{Bell_CH}, based on an extension of quantum eraser experiments\cite{Scully_Druhl,Scully_ES_a,Scully_ES_b,Scully_exp,Walborn}. 
In addition to reemphasizing the need to develop a logically consistent theory, this idealized example also provides a framework in which the relationship between collapse and (other) elementary processes can be analyzed.

To explore this relationship, in Section 3 I briefly review Gisin's theorem\cite{Gisin_c}, which shows that any explanation of wave function collapse at the \textit{elementary level} must be nondeterministic in order to prevent 
superluminal signaling at the level of observation. Some related arguments are also noted.

In Section 4 I argue that this result opens the way to an explanation of the no-signaling principle in \textit{fundamental physical} terms, without any essential reference to intelligent observers. The possibility of nondeterministic changes in elementary systems prevents the perfect copying of unknown quantum states (as exemplified in the No-cloning theorem\cite{No_Clone}), and thus prevents the instantiation of \textit{reproducible} information in isolated microsystems. Such nondeterministic changes can also eliminate physical traces of previously 
entangled states, so, even if these changes originate in spacelike-separated regions, the change cannot be recognized without local correlating interactions with another system that stores appropriate reference information. The no-signaling principle can then be defined as a prohibition of the acquisition of information about spacelike-separated occurrences by localized physical systems.

The argument is carried forward in Section 5 by noting that the principle would be violated if wave function collapse is induced by any process other than the correlating interactions by which information is acquired. This leads to the hypothesis that collapse results from (presumably small) deviations from linear evolution associated with the sort of nonlocally entangling interactions that constitute the problematic measurements.

Section 6 deals with the problem of reconciling the nonlocal and nondeterministic nature of wave function collapse with the relativistic description of spacetime. The reconciliation can be achieved by recognizing that the no-signaling principle is not just a more abstract way of expressing the limitations imposed by the speed of light. It is a genuine extension of Einstein's original postulates\cite{Orig_Rel} aimed at regulating the nonlocal quantum effects.\footnote{The principle is implemented as the Born probability rule in ordinary quantum mechanics, and as the requirement of local commutativity in quantum field theory.} This perspective leads to a new understanding of relativity - not as characterizing the causal structure of spacetime, but as describing the transformation properties of observable quantities  within the overall mathematical structure of physical theory. This allows us to incorporate a description of nonlocal quantum effects into that mathematical structure, rather than inserting 
ad hoc interpretive rules at some ill-defined point. The simplest way to construct such a description is by assuming  that nondeterministic, nonlocal effects are sequenced in some manner. By defining a global sequencing parameter,
$s$, associated with an unobservable, randomly evolving spacelike hypersurface, $\sigma(s)$, we can provide a mathematical framework in which to describe their evolution. The evolving surface remains below the threshold of observability due to the lack of complete determinism in elementary interactions.

By specializing to the case in which the parameter, $s$, coincides with the time, $t$, in some (unknown) inertial frame it is possible to adapt the substantial body of work on nonrelativistic stochastic collapse 
equations\cite{Gisin_c,Pearle_1976,Pearle_1979,Gisin_1984,Diosi_1,Diosi_2,Diosi_3,GPR,Adler_Brun} in order to formalize the hypothesis that collapse originates in nondeterministic effects associated with entangling interactions. This is done in Section 7. The stochastic modification of the Schr\"{o}dinger equation is defined in terms of the potential energy operator associated with the interactions between the particles involved. This provides an explanation of collapse at a fundamental level that parallels the standard macroscopic account of projection induced 
by measurement at the observable level, and at that level, it maintains the usual conservation laws\cite{Gillis_2}.

Because the proposed equation implies that collapse originates in incremental effects associated with the same kinds of processes that induce decoherence, it helps explain why no clear deviations from linear evolution have been observed in the recent experiments involving systems that approach mesoscopic scales\cite{Bio_OReilly,Gisin_a}, as measured by the number of elementary particles involved. The hypothesis suggests that a more relevant measure of macroscopicity is the number of entangling interactions experienced by any elementary particles in generating the entangled state. Although the number of elementary particles in these experiments is large, the number of entangling interactions undergone by any individual particle is quite small. 

The smallness of the deviations from linearity (at least in ordinary electromagnetic interactions) has also been demonstrated by a number of experiments that have exhibited the persistence of superposition in a variety of contexts\cite{Scully_Druhl,Scully_ES_a,Scully_ES_b,Scully_exp,Walborn,Vandersypen,Steffen}. Therefore, it would be helpful to have some guidance from a specific collapse proposal in trying to design a way to detect such deviations.

In Section 8 the stochastic collapse equation is used to examine the dependence of the deviations on the interaction energy that induces the entanglement, and on the relative amplitudes of interacting and noninteracting components of the wave function. The relationship derived there suggests a general strategy for potential experiments. But it is also indicates that implementing this strategy will require some considerable advances in experimental techniques.

Section 9 summarizes the argument.

\section{Idealized Test of Wave Function Collapse}
\label{sec:2}

Projection implies a violation of the principle of superposition after a finite number of interactions. It can be shown that this results in differences in some predicted measurement correlations depending on when the projection is assumed to occur. Measurements consist of entangling interactions. The key to testing for superposition effects after entangling interactions occur is to observe the 
system in a basis that is different from the basis in which the initial interactions take place. For example, one designs an experiment to correlate the x-components of spin of an elementary particle to the states of a target system, and then tests for correlations between the subject particle and the detector system in the y or z-spin basis. This is similar to what Bell demonstrated in \cite{Bell_CH}. It is the central idea in the quantum eraser experiments proposed by Scully and his 
colleagues\cite{Scully_Druhl,Scully_ES_a,Scully_ES_b,Scully_exp,Walborn}, and it is the basic ``trick" in quantum computing\cite{Mermin}. The illustration presented here combines features of Bell's example and quantum eraser arrangements.\footnote{Essentially the same example was presented earlier in \cite{Gillis_E}.}.

Consider a spin-${\frac{1}{2}}$ particle in a z-up state. If the x-up and x-down components are somehow separated and then recombined the z-up state can be recovered if the phases are properly controlled:    
\begin{equation}\label{eqno0}
|z\!\uparrow\rangle \; \Longrightarrow \; (1/\sqrt 2)(|x\!\uparrow\rangle + |x\!\downarrow\rangle \;) \Longrightarrow \; |z\!\uparrow\rangle.
\end{equation}
Suppose now that we add a ``detector" particle in an x-down state and design an interaction so that a subject particle in an x-up state flips the state of the detector from down to up, while a subject particle in an x-down state leaves the detector unchanged. The evolution of the system can be represented schematically as follows:
\begin{equation}\label{eqno1}
\begin{array}{ll} 
|z_0\!\uparrow\rangle\otimes|x_1\!\downarrow\rangle \; \Longrightarrow \;
(1/\sqrt 2)(|x_0\!\uparrow\rangle + |x_0\!\downarrow\rangle \;)
\otimes|x_1\!\downarrow\rangle  & \\   \Longrightarrow \;                      
(1/\sqrt 2)(|x_0\!\uparrow\rangle|x_1\!\uparrow\rangle + |x_0\!\downarrow\rangle|x_1\!\downarrow\rangle )
= \; (1/\sqrt 2)(|z_0\!\uparrow\rangle|z_1\!\uparrow\rangle + |z_0\!\downarrow\rangle|z_1\!\downarrow\rangle).
\end{array}
\end{equation}   
The $|z_0\!\uparrow\rangle$ state can no longer be detected in 100 per cent of the cases, but superposition effects are still exhibited through the perfect correlations between 
$|z_0\!\uparrow\rangle$  and $|z_1\!\uparrow\rangle$, and between $|z_0\!\downarrow\rangle$  
and $|z_1\!\downarrow\rangle$.

This general schema can be applied to quantum eraser experiments. The two different 
representations of the final state, 
$(1/\sqrt 2)(|x_0\!\uparrow\rangle|x_1\!\uparrow\rangle + |x_0\!\downarrow\rangle|x_1\!\downarrow\rangle )$,   and   \newline 
$(1/\sqrt 2)(|z_0\!\uparrow\rangle|z_1\!\uparrow\rangle + |z_0\!\downarrow\rangle|z_1\!\downarrow\rangle)$,  correspond to complementary 
observables. The x-representation can be thought of as containing the ``which-path" 
information, while the z-representation can be regarded as exhibiting ``interference" effects.

The important point here is that we believe that the ``interference" terms can still be seen because the elementary interaction between $|x_0\!\uparrow\rangle$ and the 
$|x_1\rangle$ ``detector" system did not result in a (complete) projection to either 
$(|x_0\!\uparrow\rangle|x_1\!\uparrow\rangle$ or 
$ |x_0\!\downarrow\rangle|x_1\!\downarrow\rangle$. 

To see this explicitly, one can expand the two x-branches separately in terms of 
the z-spin basis. 
\begin{equation}\label{eqno2} 
\begin{array}{lcl} (1/\sqrt{2})|x_0\!\uparrow\rangle|x_1\!\uparrow\rangle 
& = & (1/\sqrt{2})^3(|z_0\!\uparrow\rangle + |z_0\!\downarrow\rangle)\otimes(|z_1\!\uparrow\rangle +  |z_1\!\downarrow\rangle) \\
& = & (1/\sqrt{2})^3(|z_0\!\uparrow\rangle|z_1\!\uparrow\rangle + |z_0\!\downarrow\rangle|z_1\!\downarrow\rangle 
+ |z_0\!\uparrow\rangle|z_1\!\downarrow\rangle + |z_0\!\downarrow\rangle|z_1\!\uparrow\rangle) 
\end{array}
\end{equation}
and 
\begin{equation}\label{eqno3}
\begin{array}{lcl} (1/\sqrt{2})|x_0\!\downarrow\rangle|x_1\!\downarrow\rangle 
& = & (1/\sqrt{2})^3(|z_0\!\uparrow\rangle - |z_0\!\downarrow\rangle)\otimes(|z_1\!\uparrow\rangle - 
|z_1\!\downarrow\rangle) \\
& = & (1/\sqrt{2})^3(|z_0\!\uparrow\rangle|z_1\!\uparrow\rangle + |z_0\!\downarrow\rangle|z_1\!\downarrow\rangle 
- |z_0\!\uparrow\rangle|z_1\!\downarrow\rangle - |z_0\!\downarrow\rangle|z_1\!\uparrow\rangle)
\end{array}
\end{equation}  

The correlations in \ref{eqno1} result from the cancellation of the up-down cross terms, 
($|z_0\!\uparrow\rangle|z_1\!\downarrow\rangle$ 
and  $|z_0\!\downarrow\rangle|z_1\!\uparrow\rangle$), that occurs when \ref{eqno2} and \ref{eqno3} are added. If we were to regard the interaction between $|x_0\!\uparrow\rangle$ and the $|x_1\rangle$ system as a ``measurement", we would expect projection to one of the two x-branches with the other branch being eliminated. There would be no cancellation of cross terms and no ``interference", that is, the z-spin correlations would completely disappear.

What Bell showed is that the differences in the experimental predictions of continued superposition as opposed to projection are \textit{not} removed by adding any finite number of interactions.

Suppose that the single detector particle, $|x_1\rangle$, is replaced by $N$ 
detectors, $|x_{d1}\rangle$, $|x_{d2}\rangle$, ..., $|x_{dN}\rangle$. The subject particle is still labeled, $|x_0\rangle$, and the interactions are again designed so that an x-up subject state flips the detector particles from down to up, and an x-down state leaves them unchanged. The initial state: \newline  
$\;\;\;(1/\sqrt{2})( |x_0\!\uparrow\rangle + |x_0\!\downarrow\rangle) \otimes
(  |x_{d1}\!\downarrow\rangle ...|x_{dN}\!\downarrow\rangle ) \;\;\; $
evolves to: \newline
$(1/\sqrt{2})( |x_0\uparrow\rangle) \otimes
(  |x_{d1}\;\uparrow\rangle ...|x_{dN}\;\uparrow\rangle )
+(1/\sqrt{2})(  |x_0\;\downarrow\rangle) \otimes
(  |x_{d1}\;\downarrow\rangle ...|x_{dN}\;\downarrow\rangle ).$

\noindent Now expand the x-up and x-down branches in the z-spin basis:
\begin{equation}\label{3p8}
\begin{array}{lcl}
&   &  (1/\sqrt{2})|x_0\!\uparrow\rangle|x_{d1}\!\uparrow\rangle...|x_{dN}\!\uparrow\rangle \\
& = &  (1/\sqrt{2})^{N+2}[\,(|z_0\!\uparrow\rangle \, +  \, 
|z_0\!\downarrow\rangle)\otimes(|z_{d1}\!\uparrow\rangle \, + \, |z_{d1}\!\downarrow\rangle)...(|z_{dN}\!\uparrow\rangle \,+ \, |z_{dN}\!\downarrow\rangle)\,] 
\end{array}   
\end{equation}   
and 
\begin{equation}\label{3p9}
\begin{array}{lcl}
&   &  (1/\sqrt{2})|x_0\!\downarrow\rangle(|x_{d1}\!\downarrow\rangle...|x_{dN}\!\downarrow\rangle  \\
& = &  (1/\sqrt{2})^{N+2}[\,(|z_0\!\uparrow\rangle \, - \,  |z_0\!\downarrow\rangle)\otimes(|z_{d1}\!\uparrow\rangle \, - \, |z_{d1}\!\downarrow\rangle)...(|z_{dN}\!\uparrow\rangle \, - \, |z_{dN}\!\downarrow\rangle].    	 
\end{array}   
\end{equation}    

If $\ref{3p8}$ and $\ref{3p9}$ are expressed as sums of products of the form 
$ |z_0\rangle |z_{d1}\rangle|z_{d2}\rangle...|z_{dN}\rangle $, the terms in the two equations would be identical, but in $\ref{3p9}$ half of them would have plus signs and half would have minus signs. This makes it convenient to rewrite the two expressions as: \newline  
$  (1/\sqrt{2})^3[|z_0\!\uparrow\rangle|z_d\downarrow_{EV}\rangle  +  |z_0\downarrow\rangle|z_d\downarrow_{OD}\rangle   +  |z_0\uparrow\rangle|z_d\downarrow_{OD}\rangle  +  |z_0\downarrow\rangle|z_d\downarrow_{EV}\rangle ]$, \newline 
$ (1/\sqrt{2})^3[\,|z_0\!\uparrow\rangle|z_d\!\downarrow_{EV}\rangle \, + \, |z_0\!\downarrow\rangle|z_d\!\downarrow_{OD}\rangle 
\, - \, |z_0\!\uparrow\rangle|z_d\!\downarrow_{OD}\rangle \, - \, |z_0\!\downarrow\rangle|z_d\!\downarrow_{EV}\rangle ]$, \newline
where the following definitions have been used: \newline  
$|z_d\!\downarrow_{EV}\rangle \equiv  (1/\sqrt{2})^{N-1}\sum_{even|z_{di}\!\downarrow\rangle}{(\prod_i{|z_{di}\rangle)}}, $ \newline    
$|z_d\!\downarrow_{OD }\rangle \equiv  (1/\sqrt{2})^{N-1}\sum_{odd|z_{di}\!\downarrow\rangle}{(\prod_i{|z_{di}\rangle)}},$\newline  
where $i$ ranges from $1$ to $N.$ In other words $|z_d\!\downarrow_{EV}\rangle$ is the normalized sum of all product detector states with an even number of $|z_{di}\!\downarrow\rangle$ occurrences, and $|z_{d}\!\downarrow_{OD}\rangle$ is the corresponding sum with an odd number of $|z_{di}\!\downarrow\rangle$ occurrences.

If no projection effects have occurred after the $N$ correlating interactions in the x-spin basis, then the two branches represented by $\ref{3p8}$ and $\ref{3p9}$ can be superposed. This results in cancellation of the $\;|z_0\!\uparrow\rangle|z_d\!\downarrow_{OD}\rangle \; $ 
terms and the    
$\;|z_0\!\downarrow\rangle|z_d\!\downarrow_{EV}\rangle $ terms. So the signature 
of continued superposition is now exhibited as a perfect correlation between up results of z-spin measurements on the subject particle, and an \textit{even} number of down results from z-spin measurements on the detector particles, and a corresponding correlation between down results of z-spin measurements on the subject particle, and an \textit{odd} number of down results from z-spin measurements on the detector particles. 

So wave function collapse \textit{is observable} in principle. An experiment showing a statistically significant number of coincidences of z-up results on a subject particle and a $|z_d\!\downarrow_{OD}\rangle $ state would be strong evidence of the breakdown of superposition.

Obviously, as the number of interactions, $N$, increases it becomes more difficult to track the z-spin states of all of the detector particles. These are the typical effects of decoherence. The challenge in detecting the correlations is similar to that of constructing a quantum computer. One must carefully design the interactions that generate the entangled states, and then accurately track the evolution of all of the particles involved.

The experiments cited in the previous 
section\cite{Scully_Druhl,Scully_ES_a,Scully_ES_b,Scully_exp,Walborn,Bio_OReilly,Gisin_a,Vandersypen,Steffen} have not shown any deviations from linear evolution associated with individual interactions that exceed ordinary experimental uncertainties. If there really are deviations they might begin to reveal themselves almost anywhere between elementary and macroscopic scales. Given this huge range of uncertainty any proposed explanation of wave function collapse must be very well motivated. In the next several sections I argue that the needed motivation can be found by exploring the relationship between wave function collapse and the no-signaling principle.

\section{ No-Signaling and Wave Function Collapse \newline Imply Nondeterminism }
\label{sec:3}

The no-signaling principle is essential to the relativistic description of spacetime, and it can also be used to derive many of the key properties of quantum theory. The general reasons that the principle proves so useful have been studied by Svetlichny\cite{Svetlichny_a,Svetlichny_b}. He pointed out that nonlocal effects lead to violations of no-signaling unless they are very tightly constrained. The work of Gisin and others has shown that, in particular, the principle, in conjunction with the assumption of wave function collapse, determines the general form of both the deterministic and probabilistic types of quantum evolution. 

Gisin\cite{Gisin_c} was able to show that any deterministic quantum evolution must be linear, and that any description of projection at the elementary level must be nondeterministic. He did this by examining the ways in which density matrices can evolve. He considered distinct mixtures of quantum states such as  $\{\frac{1}{2}|x\!\uparrow\rangle;\; \frac{1}{2}|x\!\downarrow\rangle \} $, 
and $\{\frac{1}{2}|z\!\uparrow\rangle;\; \frac{1}{2}|z\!\downarrow\rangle \} $ that have the same density matrix: \newline
$ \frac{1}{2}|x\!\uparrow\rangle \langle x\!\uparrow| + \frac{1}{2}|x\!\downarrow\rangle \langle x\!\downarrow|  \; = \; 
\frac{1}{2}|z\!\uparrow\rangle \langle z\!\uparrow| + \frac{1}{2}|z\!\downarrow\rangle \langle z\!\downarrow|   
\;\;=\;\;\left[\begin{array}{cc}
1/2  &   0  \\
0   &  1/2 \\
\end{array}\right].
$
He proved that any mixed state could be obtained from a pure entangled state in a higher dimensional Hilbert space. The simple mixed states presented here can be obtained from: \newline 
$	(1/\sqrt 2)(|x_0\!\uparrow\rangle|x_1\!\uparrow\rangle + |x_0\!\downarrow\rangle|x_1\!\downarrow\rangle )
= \; (1/\sqrt 2)(|z_0\!\uparrow\rangle|z_1\!\uparrow\rangle + |z_0\!\downarrow\rangle|z_1\!\downarrow\rangle). $
Formally, the reduced density matrix representing the mixed states can be derived from the entangled state by tracing out the full density matrix over the states of one of the particles. Experimentally, the mixed states can be gotten by measuring one particle in either the $|x\!\rangle$ or $|z\!\rangle$ spin basis, and projecting out the result.

For an experimenter with access to only one of the particles, the probabilities of measurement outcomes are completely determined by the (reduced) density matrix. So, to prevent signaling, any combination of deterministic evolution and projection (brought about by operations on the other particle) must produce the same effect on the density matrix. Any dependence on the particular mixture (that is, on the specific basis states) would result in observable differences. Any nonlinearity in the deterministic evolution would generate a state-dependent difference in the density matrix. Hence, the deterministic evolution must be strictly linear, and the nonlinear projection must be nondeterministic.

Gisin's conclusion is reinforced by noting its relationship to a couple of other key results. For spacelike-separated measurements, the no-superluminal signaling principle implies the critical condition of ``noncontextuality" used in Gleason's theorem\cite{Gleason}. The theorem shows that there is a \textit{unique} way of assigning probabilities (the Born rule\cite{Born}) to measurement outcomes. Any deviation from the Born prescription (including a deterministic one) would enable superluminal signaling. Together with Gisin's result, this shows that all quantum evolution, both deterministic and probabilistic is determined by the joint assumptions of no-signaling and wave function collapse. (It is also worth noting, given all the rather strained attempts to ``derive" the Born rule in various interpretations, how easily it follows from these two assumptions.)

The relationship between the restrictions on deterministic evolution and the no-signaling principle is exhibited in a somewhat different way in the No-cloning theorem\cite{No_Clone}. The primary proof for the theorem is based on the linearity of deterministic quantum processes. However, an alternate proof involves pointing out that a perfect universal cloning process would enable superluminal signaling.  

Other important and somewhat related consequences involving the conjunction of these joint assumptions have been demonstrated by a number of authors\cite{Elitzur_1992,Popescu,Elitzur_Dolev_1,Qi-Ren,Masanes}.

Both the no-signaling principle and wave function collapse are typically regarded as descriptions about how quantum theory is manifested macroscopically, at the level of observation. Gisin's theorem is important because (among other reasons) it reveals a very important consequence of these macroscopic statements at the most fundamental level. In the sections that follow I will argue that these two assumptions jointly have further implications about how macroscopic observations are connected to elementary processes.

\section{ Nondeterministic Effects Imply \newline Limits on Information }
\label{sec:4}

Prior to the development of quantum theory the impossibility of superluminal signaling could be viewed as a fundamental limit on the speed with which \textit{any} physical process could propagate. However, the nonlocal quantum effects identified by Bell\cite{Bell_1,EPR} make it impossible to maintain this explanation for the prohibition. In ``La nouvelle cuisine" \cite{Bell_LNC} Bell pointed out that the identification of these effects meant that this central principle of contemporary theory no longer had any clear connection to fundamental physical processes. He closed his analysis with a challenge ``to couple [[the no-signaling principle\footnote{Bell referred specifically to local commutativity, which he took to be the formal expression of the no-signaling principle in 	quantum field theory.}]] with sharp internal concepts, rather than vague external ones."

Bell's challenge can be addressed by defining the no-signaling principle as a prohibition of the transmission of certain types of information. The idea is to characterize the principle in strictly physical terms (without reference to intelligent observers) in such a way that the prohibition is respected. Such a characterization carries with it the implicit assumptions that information is a property of certain kinds of (not necessarily conscious) physical systems, and that these systems are localized. The approach taken here is dependent on the result that was reviewed in the previous section, that is, that wave function collapse implies that some fundamental physical processes are nondeterministic.

The concept of information that will be employed borrows important features from that developed by Landauer and Bennett to explore the fundamental physical limits of computation\cite{Landauer,Bennett_1973,Bennett_1982,Bennett_Landauer,Landauer_size,Landauer_99,Bennett_2003}. In \cite{Landauer_99} Landauer stated ``Information is not an abstract entity but exists only through a physical representation, thus tying it to all the restrictions and possibilities of our real physical universe" and ``...information is inevitably inscribed in a physical medium". This concept must be very carefully distinguished from some different and even somewhat conflicting notions that have been used in the discussion of foundational issues over the past few decades. The viewpoint adopted here differs sharply from that which is often summarized in the slogan, ``It from bit"\cite{Wheeler_it_bit}. I do \textit{not} assume that physics is `about' information. The relevant kind of information is about the states of physical systems and about the possibility of certain kinds of physical occurrences. In order to answer Bell's challenge to make the no-signaling principle precise, it is necessary to explain how fundamental features of nature limit the ways in which this sort of information can be represented and transmitted.\footnote{This concept of information is also different from the very abstract notion that Maudlin used in 	his comprehensive study of the nonlocal nature of quantum effects\cite{Maudlin}.  Roughly speaking, he was using `information' to refer to the full set of properties that characterize a quantum state. One of his primary concerns 	was to develop a measure of the nonlocality implied by the Bell-EPR correlations, and this abstract notion is very 
	well suited for that purpose. However, it is not reproducible (as Maudlin readily acknowledges). It can also be transmitted across spacelike intervals.}

Consider, first, the necessity for physical representations of information to be \textit{reproducible}, and how this requirement is affected by the nondeterministic effects discussed above. Landauer pointed out that nondeterministic effects imply a minimal size for physical systems that instantiate information\cite{Landauer_size}. 
He assumed that the physical processes involved in artificial computational devices are effectively nondeterministic due to the presence of noise and finite limits of precision.\footnote{In Landauer's view this effective nondeterminism exists despite an underlying determinism in the dynamic evolution.} Landauer and Bennett were especially concerned with how this assumed indeterminism affected the reversibility of computational processes. They recognized that to maintain reversibility it is necessary to insure the stability of the information generated in the processes against nondeterministic effects: 
\begin{quotation}\noindent
	``It is important to understand that the immunity [[from the nondeterministic effects]] is 
	bought at the expense of the size of the potentials involved, or - more or less equivalently - 
	the size of the parts, and not at the expense of energy dissipation."\end{quotation} 
The property of reproducibility is affected in the same way as reversibility by the lack of determinism. From the perspective adopted here, the ultimate source of the indeterminism is different, but the implications for the concept of information are similar. Gisin's demonstration that wave function collapse entails a degree of indeterminism in some elementary processes implies a lower limit on the size of physical realizations of reproducible information.

Given the assumption that physical instantiations of information must be reproducible, the possibility of nondeterministic effects at the fundamental level means that information cannot be instantiated in isolated elementary systems. (The No-cloning theorem exemplifies this point.)

To illustrate the idea, consider what happens in a measurement on a single particle in an unknown spin state, using an idealized arrangement similar to that of Section 2. The experimenter chooses an axis, which can be labeled 'x', along which to measure. The $|x_0\!\uparrow\rangle$ and $|x_0\!\downarrow\rangle$ states are separated, and the x-up branch undergoes an interaction with a set of detector particles. If no projection has occurred after $N$ correlating interactions the state can be represented as: 

$\alpha\, |x_0\!\uparrow\rangle \otimes
(  |x_{d1}\!\uparrow\rangle |x_{d2}\!\uparrow\rangle ...|x_{dN}\!\uparrow\rangle )
\;+\;\beta \,  |x_0\!\downarrow\rangle \otimes
(  |x_{d1}\!\downarrow\rangle |x_{d2}\!\downarrow\rangle ...|x_{dN}\!\downarrow\rangle )$,\newline    
where $\alpha$ and $\beta$ are unknown. Even if $N$ is quite large, as long as the system remains in a superposition of $|x\!\uparrow\rangle$ and $|x\!\downarrow\rangle$ states, there is no way to take a statistical sample of the states of the detector particles since they are entangled. Any complete x-state measurement on one of them would collapse the entire system to either the $|x\!\uparrow\rangle$ or $|x\!\downarrow\rangle$ branch. Information about the relative magnitudes of $\alpha$ and $\beta$ would be lost. It is only at the stage when projection occurs that the system is transformed into a factorizable state of multiple copies of the \textit{resulting} information. So wave function collapse is an essential part of the information acquisition (and generation) process. The lack of determinism prevents reproducible information from being instantiated at scales below that at which collapse typically takes place.

It is important to recognize that wave function collapse is involved in both the creation and the destruction of information. For example, in the situation described above, if a measurement results in an $|x\!\uparrow\rangle$ outcome, then the transformation to a factorizable state eliminates any trace of the $|x\!\downarrow\rangle$ state and any physical record of the coefficient, $\beta$, within the previously entangled system. Therefore, the physical system consisting of \textit{only} those particles involved in the original entanglement relations cannot instantiate information that such a change has taken place. Such a recognition can only be made by a larger system that is capable of comparing the resulting (collapsed) state of the system with a stored representation of the initial state. Furthermore, the state representing the fact \textit{that} the collapse has occurred does not, in any way, indicate \textit{how} or \textit{where} the collapse was induced.

These considerations can be used to define the no-superluminal-signaling principle as a statement that it is impossible for a localized physical system to acquire information about a spacelike-separated occurrence, if that information is not already included in the past light cone of the system at the time that the information is acquired. This formulation does not yet guarantee that the principle is correct. However, if we assume that the principle holds then we can derive very interesting constraints on the way in which nonlocal quantum effects are induced. This issue is addressed in the next section.

\section{ No-Signaling Implies That Collapse \newline Is Induced By Entangling Interactions \newline }
\label{sec:5}

The no-signaling principle tightly constrains nonlocal effects like wave function collapse. The minimal assumption that it holds at the level of observation already allows one to infer that such effects must be nondeterministic. The main goal here is to derive specific constraints on the way in which wave function collapse is connected to elementary processes. The following heuristic argument illustrates the essential idea.

Suppose that the wave function of a particle has bifurcated into two principal components, and that a measurement apparatus has been set up to determine whether it is in one of the branches. A negative outcome indicates that the wave function has collapsed, and that the particle is localized in the other branch. This measurement clearly reveals information about the state of the (spacelike-separated) particle. If this collapse were not induced by the physical processes that constitute the measurement, then the measurement would be \textit{revealing information about a random event that occurred at spacelike separation}. The prohibition of superluminal information transmission would be violated. Therefore, collapse must be induced by some physical processes that are essential to measurement. At the most basic level, measurements establish correlations among the states of elementary particles. Therefore, the elementary interactions that generate the correlations are the source of collapse effects. 

Information that a random event such as a collapse has occurred in some location \textit{is information}. Unless the collapse is triggered by the information acquisition process then superluminal information transmission is possible.

In the situation just described, the measurement process with the negative outcome generates the collapse and acquires information about the state of the distant particle. Although it induces a nonlocal change in the state of the particle, no information about the new state exists at the distant location because the isolated particle has not interacted with any other systems. To further illustrate the point, let us modify the example and consider a pair of elementary particles in a singlet state. If the particles are labeled $a$ and $b$, the state can be represented as: 
$(1/\sqrt 2)(|x_a\!\uparrow\rangle|x_b\!\downarrow\rangle - |x_a\!\downarrow\rangle|x_b\!\uparrow\rangle )$. 
After the particles are separated, an x-spin measurement on $a$ projects it into an $|x\!\uparrow\rangle$ or $|x\!\downarrow\rangle$ state, and $b$ into the complementary state. Information about the state of $a$ is generated in the system of particles that interact with $a$. That information is reproducible and transmittable to other systems. The system of particles contains only contingent information about the state of $b$, since the information resulting from the set of interactions undergone by $a$ cannot rule out the  possibility that $b$ has interacted in some alternate basis. If $b$ does not interact in some organized 
fashion, then there is no reproducible, transmittable information in $b$, or in its immediate vicinity, since the state that would result from any such interactions is uncertain (until the information from the $a$ system is transmitted to the region of $b$). 

If measurements are made in both branches then information that a collapse has occurred exists in the apparatus of both branches, but there is no information about which measurement induced the collapse.

It is important to again emphasize that although the example described above involves a very idealized  and highly organized system, the intervention of intelligent observers is  \textit{not} essential to the applicability of the concept. Information, in the sense intended here, can be instantiated in naturally occurring systems. Perhaps, the most common example is the generation of information about the position of a particle. If the wave function of a particle separates into different branches or simply diffuses over a large enough region, it will eventually interact with systems consisting of a sufficient  number of particles to  relocalize it (either within the region of interaction or outside it). Other natural  ``information processing  devices" might include biological systems. It is plausible to suggest that  basic metabolic processes might be capable of completely reducing the wave function.

The main point can be summarized as follows. If a nondeterministic change takes place in a localized system of particles, information about that change can only be acquired by interrogating that system through an appropriate correlating interaction. If it is just such interactions that potentially induce these nondeterministic effects, then the resulting state of the system acquiring the relevant information is the same whether the change was brought about locally or by some more distant interaction. In this way signaling across spacelike intervals is prevented.

As stated above, the overall goal is to better understand how wave function collapse is connected to elementary processes. At this point we can see the general outlines of this connection. Note first that simple verification measurements, in which  the measured system is already in an eigenstate of the measured observable, do not result in projection.  The problematic cases are those in which measurement interactions \textit{nonlocally entangle} the states of the subject and detector systems. So let us hypothesize that nonlocal, nondeterministic projection effects are induced by elementary interactions that create  entanglement relations between particles that are spacelike-separated. 

The general nature of these effects can be discerned by considering what happens in measurements at the elementary level. The particles making up the detector either interact or fail to interact with the subject particle,\footnote{The interactions can be indirect through a chain of mediating particles.} indicating the subject particle's presence or absence. The wave function of the particle changes, reflecting the transfer of amplitude either into or out of the interacting branch. This suggests that elementary interactions that generate entanglement also involve a transfer of amplitude either into or out of the interacting component of the wave 
function.\footnote{This assumes that the relevant system has both interacting and noninteracting components. In entanglement-generating interactions such as parametric down conversion, there is no noninteracting component 	to be involved in any amplitude transfer.}

The binary decomposition of the wave function into interacting and noninteracting branches defines the collapse basis. (The distinction between ``interacting" and ``noninteracting" branches will be made more precise in Section 7, in which a stochastic collapse equation is described.) The idealized situation described in Section 2 can be used to describe some general features of the collapse process. In that example, after $N$ correlating interactions the state was represented as:  \newline  
$   \;\;\alpha \; |x_0\!\uparrow\rangle \otimes
|x_{d1}\!\uparrow\rangle |x_{d2}\!\uparrow\rangle ...|x_{dN}\!\uparrow\rangle 
\;+ \;\;\beta \; |x_0\!\downarrow\rangle \otimes
|x_{d1}\!\downarrow\rangle |x_{d2}\!\downarrow\rangle ...|x_{dN}\!\downarrow\rangle.\;\;\;\;\;\;\;\;\;\;$ \newline
The coefficients of the two branches given in Section 2, $(1/\sqrt{2})$, have been replaced here by $\alpha$ and $\beta$ because they are now being considered as dynamic entities which can change in response to the interactions (assumed to be taking place in the $|x_i\!\uparrow\rangle$ branch).  For example, if $|x_{dN}\!\uparrow\rangle$ interacts with the $|x_{dN+1}\rangle$ particle, then there is some probability that $\alpha$ and $\beta$ will change to $\alpha'$ and $\beta'$. The idea is that with enough such random transfers (or a large enough transfer) the amplitudes, $\alpha$ and $\beta$, will change to $1$ and $0$ or to $0$ and $1$.

The possible amplitude transfers must reproduce the Born rule exactly at the macroscopic level, in order to prevent superluminal signaling and preserve relativity. At this stage of the analysis we can surmise that each elementary interaction carries some probability, $p$, of an amplitude transfer either into or out of the the interacting component of the wave function. We can define 
$p = q + r$, where $q$ is the probability of an amplitude transfer into the interacting component, and $r$ is the probability of a transfer in the opposite direction. Since the Born rule is expressed in terms of the absolute square of the amplitude it is convenient to represent a possible increase or decrease in the amplitude of the interacting component as:
$\;\;\; \alpha'\alpha'^* = \alpha\alpha^* + d,\;\;\;$ and $\;\;\;\beta'\beta'^* = \beta\beta^* - d,\;\;\;$ 
or $\;\;\; \alpha'\alpha'^* = \alpha\alpha^* - e,\;\;\;$ and $\;\;\;\beta'\beta'^* = \beta\beta^* + e.\;\;\;$ 
To insure consistency with the rule, the expectation value of a change in the absolute square of the amplitude must be zero. So the probability of an increase times the potential increase must equal the probability of a decrease times the potential decrease: \newline  $ q * d = r * e$.

The most straightforward way to implement this general approach is to assume that $p = 1$, $q= r$, and $d=e.$\footnote{This formulation still allows  $d=e$ to vary throughout the collapse process.} The experiments cited in the introduction showing the persistence of superposition after a few interactions indicate that the typical increments in squared amplitude are small. So the sequence of steps resulting in 
collapse can be viewed as a random walk.\footnote{Such a process was described in an earlier work.\cite{Gillis_E}} The fact that the Schr\"{o}dinger equation is linear with respect to the amplitudes implies that the nonlocal transfers do not result in violations of the no-signaling principle at the observable level. However, there are implications about how to describe such a process at a more fundamental level that require a reexamination of some assumptions concerning what we really know about the nature of spacetime. In the idealized example described above the collapse process is straightforward because the sequence of steps in the random walk is clearly determined. However, the standard relativistic temporal ordering does not provide a way to sequence the interactions that make up spacelike-separated measurements such as those that occur in EPR-Bell type experiments. In general, it is very difficult to construct a coherent account of effects that are both nonlocal and nondeterministic without assuming some underlying sequence. In order to accommodate these effects the relationship between relativity and quantum theory needs to be reexamined.\footnote{The general issues raised by attempts to reconcile nonlocal effects with 	a strictly relativistic spacetime structure have been examined in depth by Maudlin\cite{Maudlin}. D\"{u}rr et al. have discussed the matter in connection with Bohmian mechanics\cite{Bohm_Rel}.}

\section{Relativity and Nonlocal Effects}
\label{sec:6}

The problems in understanding the relationship between quantum theory and relativity stem largely from the fact that views about the relativistic structure of space and time were ``frozen in" before quantum theory was fully developed. This 
has led to the peculiar situation in which many commentators very readily question the reality of elementary particles and processes, but appear to adhere rigidly to a belief that the pre-quantum version of relativity is the last word in spacetime ontology. This sort of viewpoint runs completely counter to the \textit{methodology} that Einstein used in developing the theory. He strongly emphasized the need to base the concepts used to describe space and time on observations of physical processes. Consider the following excerpts from his 1905 paper\cite{Orig_Rel}:
\begin{quotation}\noindent 
	``... since the assertions of any such theory have to do with the relationships 
	between rigid bodies (systems of co-ordinates), clocks, and electromagnetic 
	processes." \end{quotation}

\begin{quotation}\noindent``Now we must bear carefully in mind that a mathematical 
	description of this kind has no physical meaning unless we are quite clear as to what we understand by time." \end{quotation}

We need to recognize both that observations are macroscopic (or at least mesoscopic) physical processes that involve the acquisition of stable information, and also that the observational basis for relativity has expanded since Einstein first formulated the theory. Given the experimental evidence available to him, Einstein's original derivation was based on two postulates: (1) the laws of physics are the same in all inertial frames; (2) the speed of light is constant, independent of the emitting source. In a strictly classical setting these postulates led to an understanding of relativity as a theory about the fundamental structure of space and time. The apparently natural exclusion of nonlocal effects (action-at-a-distance) was viewed as a very welcome feature of this structure. However, following the development of quantum mechanics, it eventually became clear that there are real nonlocal effects. In order to prevent these effects from enabling the superluminal transmission of information, and, thereby violating 
relativistic transformation properties, contemporary physics has, in effect, added a third postulate. In ordinary quantum mechanics the Born probability rule plays this role, while in field theory it is the assumption of local commutativity that guarantees the consistency of quantum predictions with relativity. The incorporation of nonlocal effects, regulated by the additional postulate, means that special relativity can no longer be viewed simply as a theory about a background structure in which physical processes take place. Since these probabilistic processes provide the observational basis for the mathematical description of spacetime, we must acknowledge that in the logical framework of today's physics these two aspects are tightly intertwined.

It is important to understand that all three postulates are formulated in terms of concepts that can only be defined at a level at which stable information exists. References to inertial frames (in the first postulate), and to the speed of light and state of motion of emitting sources (in the second postulate) are critically dependent on complex, stable networks of physical relations. Given the nondeterministic character of some elementary processes discussed in the preceding sections, we must recognize that these postulates do not uniquely determine spacetime structure at more fundamental levels.

With this perspective and the recognition that there is a third postulate to regulate the nonlocal quantum effects, one can hypothesize additional properties of spacetime, provided that they maintain overall consistency. A minimal assumption, that keeps all three postulates intact, is that there is a randomly evolving spacelike hypersurface on which  the nondeterministic, nonlocal effects propagate. It evolves forward in time, allowing the definition of a global sequencing parameter, $s$, with successive surfaces labeled as $\sigma(s)$. No special reference frame or even foliation needs to be singled out, and most physical processes (other than the making and breaking of nonlocal entanglement relations) are still confined to the light cone. The randomly evolving surface plays two related roles. In addition to defining the the way in which the collapse effects propagate, it also defines which entanglement relations hold when those effects propagate. 

This last point is especially important in the construction of collapse models. The most serious interpretive problems arise in situations in which spacelike-separated measurements are made on an entangled system. The measurement interactions generate additional entanglement relations, and collapse effects are propagated through these relations. If collapse is a genuine physical effect, these relations must be unambiguously determined (even though unknown) at the ``time" (that is, along the hypersurface) at which the collapse occurs. If spacelike-separated interactions are generating additional nonlocal entanglement relations on an already entangled system it is essential that the interactions have a defined sequence.

To illustrate, a little more explicitly, the need for some sort of global sequencing in order to provide a coherent account of collapse,  consider the following situation. Suppose that $\psi$, $\phi$, and $\chi$ represent elementary systems, and that $\psi$ consists of two spatially separated branches, $\psi_1$, with amplitude $\alpha$, which interacts with $\phi$, and $\psi_2$, with amplitude $\beta$, which interacts with $\chi$. The states of $\phi$ and $\chi$ prior to the interactions can be labeled $\phi_0$, and $\chi_0$, so the state of the combined system prior to interaction is $ (\alpha\psi_1 + \beta\psi_2)\otimes \phi_0\otimes \chi_0$. Given appropriate, short duration interactions, the initial product state will eventually evolve to the entangled state: 
$ (\alpha\psi_1\otimes \phi_1\otimes \chi_0) + (\beta\psi_2\otimes \phi_0\otimes \chi_2)$, according to the Schr\"{o}dinger equation. If there is no change other than that due to Schr\"{o}dinger evolution we need not worry about the temporal ordering of the spacelike-separated interactions. However, collapse models posit that there \textit{are} other changes in the state. For example, in the collapse hypothesis proposed in Section 5, the first step in the stochastic evolution of the system depends on the sequencing of the interactions. If the $\psi_1$ interaction is sequenced first, it would be to \newline 
$[(\alpha\pm \epsilon_1)\psi_1\otimes \phi_1\otimes \chi_0] + [(\beta\mp \epsilon_2)\psi_2\otimes \phi_0\otimes \chi_0]$;
otherwise, it would be to  \newline 
$  [(\alpha\mp \epsilon_1')\psi_1\otimes \phi_0\otimes \chi_0] 
+ [(\beta\pm \epsilon_2')\psi_2\otimes \phi_0\otimes \chi_2],$  where the $\epsilon$ terms represent the incremental changes in amplitude associated with the interactions. These are distinct states, and if we are to regard wave function collapse as an objective process, then the states to which the system collapses should be regarded as physically real, even if the nondeterministic nature of the process makes it impossible to track those states at every stage.

One cannot avoid the need for sequencing of spacelike-separated interactions by assuming that they occur at the same stage of stochastic evolution. In general, the interactions of $\psi_1$ and $\psi_2$ can be parts of large sets of interactions with $ \phi_a, \phi_b, \phi_c ... $ and $ \chi_a, \chi_b, \chi_c ... $ like those that take place in measurements. The two sets of interactions can be completely spacelike-separated relative to each other, but the temporal and spatial relations within the sets will be very complex. There is no reasonable way to sort out the ordering of all the interactions except with a global sequencing parameter. 

The fact that the hypersurface evolves randomly means that most spacelike-separated interactions are effectively independent since they are associated with different values of the sequencing parameter, $s$. With a large number of interactions being sequenced differently, there is an  unambiguous decomposition of the wave function into interacting and noninteracting branches at each stage of the collapse process.\footnote{In those cases in which 	interactions in spacelike-separated regions occur at the same value of $s$, the decomposition is trivial, and there 
	is no change in the amplitudes, $\alpha$ and $\beta$. Since real measurements consist of an enormous number of 	interactions, this possibility does not preclude a definite outcome.} If the interactions are parts of spacelike-separated measurements of the position of $\psi$ (or some correlated property), then the separate sequencing and effective  independence insures that one measurement will have a positive outcome, and the other will have a negative outcome.

The sequencing parameter, $s$, is similar in many respects to a preferred time coordinate such as occurs in de Broglie-Bohm theory although it does not necessarily coincide with the time in any single inertial reference frame.\footnote{Somewhat similar constructions have been developed for pilot wave theories in order to bring 	them more in line with the spirit of relativity. See \cite{Bohm_Rel}.} At each point, $x$, on $\sigma(s)$ one can define a proper time derivative, $d\tau(x)/ds$, but, in general,  $d\tau(x)/ds \neq d\tau(x')/ds$ for 
$ x \neq x'$. If $d\tau(x)/ds  > 0$ at every point, $x$, for all values of $s$, then the succession of hypersurfaces will constitute a foliation of spacetime, but this is not a necessary condition. As long as the surface remains everywhere spacelike, one could have $d\tau(x)/ds = 0$ for some values of $s$ in some regions of $\sigma(s)$. In these cases the family of surfaces that sweeps out spacetime would be more general than the set of foliations. This greater generality could be useful in some situations. These sorts of more general constructions have been considered elsewhere\cite{Gillis_E,Struyve_Tumulka}.

Under this hypothesis, the state vector evolves probabilistically and continuously with respect to $s$. At any given stage, the division of the wave function into interacting and noninteracting branches defines a binary decomposition of the Hilbert space of the system into two orthogonal subspaces.\footnote{In simple, 
	idealized cases this decomposition remains the same throughout the measurement, but, in general, it can change due to changes in the interactions that occur or due to the random evolution of $\sigma(s)$.} The precise nature of this decomposition will be described in the next section. When the state vector reaches one of the two subspaces, the distinction between interacting and noninteracting components disappears, and the decomposition becomes trivial. The probabilistic portion of the  evolution ceases until some new binary decomposition is defined by subsequent entangling interactions.

Since this approach assumes that there is an evolving spacelike hypersurface, $\sigma(s)$, that determines the actual sequencing of the nonlocal effects, the description of the evolution at this level is not Lorentz covariant. The sequencing remains unobservable, in principle, because the amplitude shifts are nondeterministic. The observed outcome of the projection process is consistent with \textit{any} sequence of spacelike-separated interactions. Any attempt to ``watch" the collapse process necessarily involves additional interactions on an entangled system that are subject to the same lack of determinism and unobservable sequencing.

The fundamentally probabilistic nature of the amplitude transfers precludes both a stable physical representation of information \textit{and} the definition of reference frames at the most elementary level. The application of these concepts requires a well-defined network of physical relations, and such a network can only exist at a scale at which probabilistic fluctuations are small compared to relevant parameters. The prohibition of superluminal information transmission and the relativistic description of spacetime emerge together at the level at which reproducible, transmittable information can be defined.\footnote{Because the definition of reference frames depends on the same nondeterministic processes that generate the nonlocal quantum effects, no finite speed can be attributed to the nonlocal 	effects in any reference frame. This is how the hypothesis of an incremental collapse process can be reconciled with the recent demonstration by Bancal, et al., \cite{Bancal_Gisin,Gisin_f} that no influences propagating at any finite speed (even a superluminal speed) can explain nonlocal quantum correlations.}

Of course, any suggestion that spacetime possesses a structure beyond what is determined by the relativistic metric will be met with extreme skepticism. Even Bell, who was very sympathetic to the de Broglie-Bohm approach, had serious 
reservations about its dependence on a preferred foliation\cite{Bell_LI}:
\begin{quotation}\noindent
	``For me, this is an incredible position to take - I think that it is quite logically consistent, 	but when one sees the power of the hypothesis of Lorentz invariance in modern physics, I think you just can't believe in it."\end{quotation}

In response to this concern one might point out that the power of the hypothesis does not require that Lorentz symmetry follows simply from the \textit{ontological structure} of spacetime. Consider the central role played in contemporary theory by gauge symmetry. The mathematical structures in gauge theories are not usually endowed with much ontological significance. Rather they are taken as characterizing the range of choices that we have in describing various physical situations. A similar freedom in describing the sequencing of spacelike-separated events associated with nonlocal quantum effects arises because our ability to observe that sequencing is limited by the nondeterministic nature of those effects. Thus, the Lorentz invariance that is usually attributed to spacetime structure can instead be understood as a property of the mathematical transformations of the \textit{description} of spacetime.

Another reason for the extreme reluctance to abandon the classical relativistic spacetime ontology is that it beautifully captures our intuitive belief that causal processes must be continuous in space and time. To alleviate discomfort on this point we need to look at the reasons for those deep-seated intuitions about causality. All of our direct experience of the world is macroscopic, and there is almost nothing in that experience that suggests that there are physical connections that are not mediated by spatial contiguity. We have become aware of another type of physical connectedness, quantum entanglement, only within the last century. But now that we have begun to explore its implications, we must broaden our perspective. Lorentz invariance can still be seen as a consequence of basic properties of nature, but now those properties must include the probabilistic nature of fundamental interactions.

As stated above, in this approach the collapse of the state vector is described  as a continuous process \textit{in Hilbert space} with respect to the sequencing parameter, $s$. The overall aim is to develop a method of averaging over the possible families of evolving spacelike hypersurfaces and over the possible sequences that are consistent with observed outcomes in order to recover a Lorentz covariant account of the evolution of physical systems at the level of observation. A first step toward this goal can be taken by considering the simplest special case of a randomly evolving hypersurface, namely, a preferred reference frame. This allows us to build on the substantial body of work dealing with nonrelativistic 
time-continuous stochastic collapse equations.

\section{A Stochastic Collapse Equation Based on \newline Entangling Interactions }
\label{sec:7}

In the work in which he showed that any description of projection at the elementary level must be  nondeterministic Gisin\cite{Gisin_c} went on to develop a time-continuous stochastic generalization of the Schr\"{o}dinger equation that results in wave function collapse. This development followed earlier works by Pearle\cite{Pearle_1976,Pearle_1979}, Gisin\cite{Gisin_1984}, and Diosi\cite{Diosi_1,Diosi_2,Diosi_3}, and provided a continuously evolving analogue of the collapse model proposed by Ghirardi, Rimini, and Weber\cite{GRW,GPR}. A comprehensive review of this body of work by Adler and Brun\cite{Adler_Brun} examined the general characteristics of norm-preserving stochastic collapse equations. The general framework developed by these various authors will be used here to construct an equation based on the idea that wave function collapse is intrinsically connected to entangling interactions.

There is also a substantial literature concerned with the problem of constructing a collapse equation in relativistic spacetime\cite{Pearle_2,Bedingham,Tumulka,Pearle_RDCM}. Based on the discussion in the preceding section the approach here is different. I assume that there is a global sequencing parameter, $s$, that labels an evolving spacelike hypersurface. For the purposes of this section, a further special assumption is made that the parameter, $s$, coincides with the time, $t$, in some unknown inertial frame.  It is important to emphasize that this time coordinate does \textit{not}, in general, coincide with the time coordinate in the laboratory frame, or any other frame in which one might naturally describe a given physical situation. This helps to insure, with high probability, that spatially separated interactions or measurements are independent. To guarantee almost complete independence one can allow a very slight ``warping" of the spacelike surface corresponding to the time coordinate since one can certainly tolerate some variations of about $10^{-18}$ seconds over atomic distances.\footnote{This slight warping does not hinder our ability to apply a nonrelativistic stochastic collapse equation since there is always some limit to the precision with which time can be measured. Very low relative atomic velocities of the order $ v/c \, \approx \, 10^{-8}$ (corresponding to temperatures of about $1^o$ K) can induce relativistic variations in time of about $10^{-18}$ seconds over atomic distances. Since nonrelativistic quantum theory works at speeds much greater than this, there is no need to assume that the spacelike surface must be perfectly ``flat".} Since stochastic fluctuations are assumed to occur continuously, even this tiny finite difference in time would make interactions separated by such distances effectively independent. The effective independence of spatially separated interactions insures compatibility with a relativistic description, while the global nature of the  stochastic process guarantees consistency among competing collapse centers.

Stochastic collapse equations can be represented in the general form:  
\begin{equation}\label{eqno 7p1}
|d\psi\,\rangle \,   = \, (-i/\hbar)\hat{H}|\psi\,\rangle dt \, 
- \, \frac{1}{2}\sum_k\hat{ \mathcal B}_k^\dagger  \hat{ \mathcal B}_k   
|\psi\,\rangle dt \,  +\, \sum_k\hat{ \mathcal B}_k|\psi\,\rangle d\xi_k, 
\end{equation} 
where $\hat{H}$ is the Hamiltonian and $(-i/\hbar)\hat{H}|\psi\,\rangle dt$ describes the ordinary Schr\"{o}dinger evolution, the $\hat{ \mathcal B_k}$ are operators that generate the stochastic modification,  and the $d\xi_k $ are independent complex Wiener processes with vanishing ensemble averages  ($M[d\xi_k ] = 0$), that obey the It$\hat{o}$ stochastic calculus: 
$d\xi^{*}_{j} d\xi_k = dt \delta_{jk}, \;\;\; dt d\xi_k = 0$. This is a slight variation of one of the formulations presented by Adler and Brun\cite{Adler_Brun}.
They show that in order to preserve the norm the $\hat{ \mathcal B_k}$ must satisfy  
$\langle \psi | \hat{ \mathcal B_k}|\psi\,\rangle \, = \, 0,$ and that, therefore, they take the form,  
$ \hat{ \mathcal B_k} \;= \;   \hat{\mathcal L_k}  - \langle \hat{\mathcal L_k} \rangle,$ where $\hat{\mathcal L_k}$ are Lindblad operators, and $ \langle \hat{\mathcal L_k} \rangle $  is the expectation value of 
$  \hat{\mathcal L_k}  $ in the state, $|\psi\,\rangle $, $\langle \psi | \hat{\mathcal L_k}|\psi\,\rangle$. (This makes clear the nonlinear dependence of the stochastic term on the particular state.)

To get an intuitive understanding of the general collapse equation, \ref{eqno 7p1},  consider a simple situation with just one collapse operator, $ \hat{ \mathcal B} $, and two possible measurement outcomes. The Lindblad operator used to construct 
$\hat{ \mathcal B}$ would be a typical (self-adjoint) observable with eigenvectors corresponding to the two possible outcomes. The ordinary Schr\"{o}dinger evolution, described by the first term in the equation, would lead to a transition such as:  
$(\alpha|S_1\!\rangle + \beta|S_2\!\rangle) \otimes|M_0\!\rangle  \Longrightarrow \;  
(\alpha|S_1\!\rangle|M_1\!\rangle + \beta|S_2\!\rangle |M_2\!\rangle,  $
where the states of the system and apparatus are indicated by  $|S_i\!\rangle $ and  $|M_i\!\rangle$, respectively. What the collapse equation must do is to produce the alternate transitions:   
$(\alpha|S_1\!\rangle + \beta|S_2\!\rangle) \otimes|M_0\!\rangle  \Longrightarrow \;   |S_1\!\rangle|M_1\!\rangle,  $
and 
$\alpha|S_1\!\rangle + \beta|S_2\!\rangle) \otimes|M_0\!\rangle  \Longrightarrow \; |S_2\!\rangle |M_2\!\rangle,  $ with the probabilities, $\alpha\alpha^*$ and $\beta\beta^*.$ Given the condition, $\langle \psi | \hat{ \mathcal B_k}|\psi\,\rangle \, = \, 0,$ it is clear that the factor in the third term in the equation, 
$ \hat{ \mathcal B}|\psi\,\rangle, $ is orthogonal to $\psi$. Heuristically, one can picture the state vector as stochastically traversing an arc between the two possible measurement outcomes with "angular" increments that are proportional to the length of 
$ \hat{ \mathcal B}|\psi\,\rangle $. The middle term in the equation, 
$ \,- \, \frac{1}{2} \hat{ \mathcal B}^\dagger \hat{ \mathcal B}|\psi\,\rangle dt \,$, is a first-order correction to compensate for the slight increase in length brought about by the stochastic action; it maintains the normalization of the state vector. The state, $\psi$, takes a random walk along the arc, and as it approaches one of the end-state eigenvectors,  $\psi_1$ or  $\psi_2$, $ \hat{\mathcal O}|\psi\,\rangle $ approaches $ \langle \hat{\mathcal O} \rangle |\psi\,\rangle $, and, hence, one can see from the definition of $\hat{ \mathcal B}$  in terms of the observable, $\hat{ \mathcal O}$, ($\hat{ \mathcal B} \;= \;   \hat{\mathcal O}  - \langle \hat{\mathcal O} \rangle,$) that $ \hat{\mathcal B}|\psi\,\rangle $ vanishes. This illustrates the way in which the stochastic action brings about convergence to one of the two expected outcomes. The variation in the length of $ \hat{\mathcal B}|\psi\,\rangle $ during the stochastic evolution is also responsible for insuring compliance with the Born probability rule. The way in which the magnitude varies implies that, at any stage of the evolution, for the wave function, $ \psi \; = \; \alpha|S_1\!\rangle |M_1\!\rangle\, + \, \beta|S_2\!\rangle |M_2\!\rangle, $  the conditional expectation value is  $\alpha\alpha^*$ for the outcome, $|S_1\!\rangle |M_1\!\rangle $, and $\beta\beta^*$ for $|S_2\!\rangle |M_2\!\rangle $.\footnote{Adler and Brun give a general proof that the collapse probabilities correspond to  $\alpha\alpha^*$ and $\beta\beta^*$, but their proof is based on the assumption that the collapse operator,  $\hat{ \mathcal B}$, is constructed from a self-adjoint operator, $\hat{ \mathcal O}$, that commutes with the Hamiltonian. That assumption does not apply to the collapse operator that will be defined below.}

What distinguishes various approaches to constructing collapse equations is the choice of the observable, $\hat{\mathcal O}$, that is used to build the stochastic operator, $\hat{\mathcal B}$. This choice determines the basis into which the state vector collapses. The two most common approaches use either the position operator or the Hamiltonian, so that the wave function is reduced either to an approximate position state or to an energy eigenstate.  Since the essential idea of this proposal is that collapse is induced by entangling interactions, the collapse operators employed here will be based on the potential energy functions that describe the interactions that establish entanglement relations between particles.

Potential energy functions are, of course, self-adjoint operators, but they differ sharply from the usual observables in that they do not have easily defined eigenvectors. Nevertheless, in measurement-like situations they do have the effect of placing the system-apparatus combination into an entangled state with well-defined orthogonal branches. It is this property that is crucial in constructing a stochastic collapse equation, and, in fact, in virtually every attempt to explain the measurement process.  

Consider that the standard von Neumann account of measurement\cite{von_Neumann} starts with this sort of entangling interaction. Decoherence accounts such as that presented by Zurek\cite{Zurek_Decoh} rely on precisely this sort of coupling, and the same is true of Everett's relative state interpretation\cite{Everett}. Pilot wave theories also, at least implicitly, assume that that it is these sorts of interactions that clearly separate the wave function components in configuration space and guide the particles to clearly defined outcomes. In virtually all the stochastic collapse proposals cited above the initial coupling of the ``system" to the ``apparatus" is brought about by an interaction potential. A key feature of the current approach is that it assigns to the measurement-like interactions the \textit{central} role in inducing collapse, rather than a major supporting role; it also emphasizes collapse to an entangled branch of a wave function involving many elementary systems, rather than to an eigenstate of a single system. The collapse equation based on interaction potentials will now be presented, followed by a discussion of how it captures the essential ideas described in earlier sections.

To simplify the construction I will limit considerations here to two-component interactions, where ``component" can refer to either elementary particles or to a subsystem of the macroscopic apparatus of any size, provided that it can reasonably be regarded as acting as a unit during the measurement process. So we can consider Hamiltonians of the form: 
$ (-\hbar^2/2m_i)\nabla_i^2 + (-\hbar^2/2m_j)\nabla_j^2 + V(x_i - x_j). $ 
In what follows, the interaction between the $i^{th}$ and $j^{th}$ components, $V(x_i - x_j)$, will be abbreviated as $V_{ij}$.

Since it is \textit{entangling} interactions that are assumed to be the source of collapse effects, it is also desirable that the collapse operator reflect the amount of entanglement that is generated by the interaction. This depends on the extent to which it alters the states of the component systems from what they would have been in the absence of the interaction. This, in turn, depends on the ratio of the interaction energy to the masses of the systems involved; the effect of the interaction tends to vary inversely with the masses. This suggests that  $ \hat{ \mathcal B} $ should be constructed from the operator, $\hat{\mathcal{V}}' \, \equiv  \,\sum_{ij} V_{ij}/(m_i+m_j)$, where the sum is over all systems in the entangled state. One must also insure that the collapse equation, \newline 
$ |d\psi\,\rangle \,   = \, (-i/\hbar)\hat{H}|\psi\,\rangle dt \, 
- \, \frac{1}{2}\sum_k\hat{ \mathcal B}_k^\dagger  \hat{ \mathcal B}_k   
|\psi\,\rangle dt \,  +\, \sum_k\hat{ \mathcal B}_k|\psi\,\rangle d\xi_k $, 
is dimensionally consistent. This requires that the term, 
$\hat{ \mathcal B}_k|\psi\,\rangle d\xi_k $, be dimensionless. In general, the Wiener process, $d\xi $, has the dimension $\sqrt{t}$. So, let us define 
$ \hat{\mathcal{V}} \,=\, \hat{\mathcal{V}}'/(c^2\sqrt{\tau_o})$, where $\tau_o$ is a time parameter related to the collapse rate, and $c$ is a  speed parameter. Although this formulation is nonrelativistic it is convenient to take $c$ as the speed of light. We can now define the collapse operator, $\hat{ \mathcal B}$, as $ \hat{\mathcal{V}} - < \hat{\mathcal{V}} >. $

This choice for the collapse operator generates an evolution of the state vector that closely parallels the qualitative discussion presented in earlier sections. At any particular stage, the stochastic action defined by $\hat{ \mathcal B} =  
\hat{\mathcal{V}} - < \hat{\mathcal{V}} >   $  separates the various branches of an entangled wave function into two components depending on whether they are experiencing an interactive potential energy above or below the average for the systems making up the wave function. As described in section 5, the stochastic action  shifts amplitude between the two components. 

In measurement situations in which the wave function has separated into several well-defined branches, one branch might be involved in measurement-like interactions while the others are not. In these situations the rough distinction between ``interacting" and ``noninteracting" components can be seen as corresponding to those in regions of configuration space where the interaction potentials show marked variation versus those in regions with little or no variation in $V_{ij}$. In more general scenarios the analysis is somewhat more involved. There could be a larger number of segments of the wave function undergoing interactions. Although the operator, $\hat{ \mathcal B}$, might have a similar effect on these segments at any particular stage, the essential independence of spatially separated collapse processes described earlier allows them to compete in a way that eventually reaches a determinate outcome. The possibility that both attractive (negative) and repulsive (positive) potential energy  functions could be involved at various points introduces an extra degree of complexity to the process, but it is not a serious impediment to reaching an outcome. Again, the effective independence of separated components, along with the fact that either attractive or repulsive interactions will tend to dominate at some stage pushes the evolution towards a specific result.  

The previous points assume that measurement effects are reasonably localized. This follows from the fact that the interaction potentials are limited by distance. In this respect the current proposal is similar to the Continuous Spontaneous Localization (CSL) approach championed by Pearle, Ghirardi, Rimini, and Bassi\cite{GPR,Bassi_Ghi}.\footnote{Pilot wave theories also emphasize the primacy of (approximate) position measurements.}  A key difference, however, is that the energy required to localize the particle is provided by the measurement apparatus, rather than by some additional, unobservable field. Also, it is the particular arrangement of interactions (characteristic of the type of measurement) that defines the local region, rather than some predetermined limit. The resulting natural delineation of distinct local regions is what  makes it possible to describe competing collapse processes (as would occur, for example, between different regions of a receptor screen in a double slit experiment.)

As noted earlier, the approach adopted here is somewhat more general than in most of the stochastic literature since it emphasizes collapse to an entangled branch of a multi-system wave function, rather than to an eigenstate of a single system. In particular, this proposal does \textit{not} imply that collapse eliminates all entanglement relations of the measured system with other systems. This residual  entanglement, along with the definition of the collapse operator in terms of interaction potentials has a very interesting consequence. Under this description of the measurement process it turns out that conservation laws are respected, not just on average, but in individual instances of wave function collapse. The interaction potentials are assumed to be conservative. The exchange of energy, momentum, and angular momentum between systems is mediated by potentials. The branches of the systems that are involved in the exchange are enhanced or suppressed together by the stochastic term, so any gain (or loss) of the relevant quantities by one system is matched by a corresponding loss (or gain) by the other. This strict conservation might be seen as an advantage for this approach.

It might seem, at first, that there is a possibility of a noticeable violation of, for example, energy conservation if the interacting and noninteracting branches of a wave function have substantially different kinetic energies. But more careful analysis shows that the variation of conserved quantities among different well-defined branches is attributable to earlier entangling interactions with systems that are also involved in the collapse of the state vector. When the value of the relevant quantities is calculated for all of the systems involved, it can be seen that the conservation laws are respected. A rather artificial example can illustrate this point. Suppose that up and down spin states of a particle are separated in the first stage of a Stern-Gerlach apparatus, and that one of them is then accelerated while the other is decelerated. If the spin-up (accelerated) branch is subsequently detected, then the collapse of the wave function also enhances those components of the accelerating apparatus that provided the additional energy to the up branch. In the same way it eliminates the components of the decelerating apparatus that removed energy from the spin-down branch. Although this example is somewhat artificial it should be clear that whatever systems were originally responsible for the difference in energy will also be involved in the collapse. This issue is discussed at greater length in \cite{Gillis_2}.\footnote{A collapse model that conserves energy on an ensemble level has been proposed by Gao in ch. 8 of \cite{Gao_1}; see also \cite{Gao_2}.}

The next section deals with the consequences of this proposal for the design of possible experiments.

\section{Assessing Experimental Prospects }
\label{sec:8}

Generically, the assumption that wave function collapse is a real phenomenon implies that there are deviations from the correlations that are predicted based on strictly linear evolution. The general types of these deviations were reviewed in Section 2. The collapse equation proposed in the previous section allows us to derive an approximate expression relating the expected deviations to the amplitudes of the interacting and noninteracting components of the wave function, and to the interaction potential. This expression can help guide the design of possible experiments. However, even with this guidance, carrying out such experiments would be very challenging.

The discussion in Section 2 focussed on the violations of the principle of superposition that would result from a collapse of the wave function after a possibly large number of correlating interactions. The proposed description of collapse as a continuous process enables us to examine the possibility of small deviations after a single entangling interaction. By limiting the number of elementary systems involved the task of looking for correlations (and possible deviations from them) is substantially simplified.

Recall the perfect correlations in two different bases that would result from the operation of an ideal quantum eraser as illustrated in equation \ref{eqno1} from Section 2:  \newline 
$\;\;\;\;\;\;	|z_0\!\uparrow\rangle\otimes|x_1\!\downarrow\rangle  \; \Longrightarrow \;
(1/\sqrt 2)(|x_0\!\uparrow\rangle + |x_0\!\downarrow\rangle \;)
\otimes|x_1\!\downarrow\rangle   \\   \Longrightarrow \;                      
(1/\sqrt 2)(|x_0\!\uparrow\rangle|x_1\!\uparrow\rangle + |x_0\!\downarrow\rangle|x_1\!\downarrow\rangle )
= \; (1/\sqrt 2)(|z_0\!\uparrow\rangle|z_1\!\uparrow\rangle + |z_0\!\downarrow\rangle|z_1\!\downarrow\rangle). $ 
\newline   The correlations are presumed to have been established by an interaction between the $|x_0\!\uparrow\rangle$ branch of the ``system" particle and the $|x_1\rangle$ ``detector" particle. For convenience let us switch to a more generic notation:
\begin{equation}\label{eqno8p1}
\begin{array}{ll} 
(1/\sqrt 2)(|S_I\!\rangle + |S_N\!\rangle \;) \otimes|D\!\rangle  & \\   \Longrightarrow \;  
(1/\sqrt 2)(|S_I\!\rangle|D_I\!\rangle + |S_N\!\rangle |D_N\!\rangle  )
= \; (1/\sqrt 2)(|S_Z\!\rangle|D_Z\!\rangle + |S_A\!\rangle |D_A\!\rangle  ).
\end{array}
\end{equation}   
In this expression, $S$ and $D$ refer to the ``system" and ``detector" particles; subscripts, $I$ and $N$, refer to interacting and noninteracting components of the wave functions, and subscripts, $Z$ and $ A$, refer to symmetric and antisymmetric superpositions of the interacting and noninteracting components (corresponding to 
the z-state representation): 
$|S_Z\!\rangle  =  (1/\sqrt 2)(|S_I\!\rangle + S_N\!\rangle \;)$   
and  $ |S_A\!\rangle  =  (1/\sqrt 2)(|S_I\!\rangle - S_N\!\rangle \;)$ (with analogous expressions for the detector particle).

Since the ``measurement" interaction was made in the  $I,N $ basis,  one expects the perfect correlation to be maintained in this basis whether or not there are any collapse effects. But, the correlation in the symmetric-antisymmetric ($Z,A$) basis depends on the assumption that the evolution is strictly linear. The perfect correlation in that basis also depends on the assumption that the two branches of the wave function are exactly equal in magnitude. Since we want to consider the possibility that the amplitudes of different components can change, let us rewrite the expression in a more general form:
\begin{equation}\label{eqno8p2}
\begin{array}{ll} 
(\alpha|S_I\!\rangle + \beta|S_N\!\rangle) \otimes|D\!\rangle  \Longrightarrow \;  
\alpha|S_I\!\rangle|D_I\!\rangle + \beta|S_N\!\rangle |D_N\!\rangle    \; \; \;  =  \\ 

(1/2)[\alpha(|S_Z\!\rangle+|S_A\!\rangle)(|D_Z\!\rangle+|D_A\!\rangle ) 
+ \beta(|S_Z\!\rangle-|S_A\!\rangle)(|D_Z\!\rangle-|D_A\!\rangle )]  \; \; \; =   \\ 

(1/2)[(\alpha+\beta)(|S_Z\!\rangle|D_Z\!\rangle + |S_A\!\rangle|D_A\!\rangle)  
+ (\alpha-\beta)(|S_Z\!\rangle|D_A\!\rangle + |S_A\!\rangle|D_Z\!\rangle )] .
\end{array}
\end{equation}                               
It is clear that if $ \alpha\;=\;\beta\;=\; 1/\sqrt 2 $ the cross terms, $|S_Z\!\rangle|D_A\!\rangle$ and  $|S_A\!\rangle|D_Z\!\rangle$, disappear, and the expression reverts to the simpler version.

If we now assume that the interaction involves a stochastic change of $\alpha$ and $\beta$ to  
$\alpha \pm \epsilon_1 $ and $\beta \mp \epsilon_2$, the last line of \ref{eqno8p2} becomes:
\begin{equation}\label{eqno8p3}
\begin{array}{ll} 
(1/2)[(\alpha+\beta \pm (\epsilon_1-\epsilon_2 ))(|S_Z\!\rangle|D_Z\!\rangle + |S_A\!\rangle|D_A\!\rangle)  
+ (\alpha-\beta \pm (\epsilon_1+\epsilon_2 ))(|S_Z\!\rangle|D_A\!\rangle + |S_A\!\rangle|D_Z\!\rangle )] .
\end{array}
\end{equation}                                
The observability of any collapse effects obviously depends on the magnitude of $\epsilon_1$ and 
$\epsilon_2$.\footnote{The reason that the changes in the amplitudes represented by $\epsilon_1$ and $\epsilon_2$ 
	are different is that the changes in the \textit{squared} amplitudes must be the same.}
The proposal of the previous section does not yield a specific numerical value, but it does imply a functional dependence of the $\epsilon$ values on the strength of the interaction, and also on the magnitudes of $\alpha$ and $\beta$. We can get an approximate expression for this dependence by making some simplifying assumptions about the potential and integrating the stochastic change through a complete interaction. Strictly speaking, the changes in $\alpha$ and $\beta$ are constant only for infinitesimal variations, but this can be ignored at this stage of approximation.

The stochastic changes are generated by the operator, 
$ \hat{ \mathcal B} \;= \;  \hat{ \mathcal E}(\hat{V} - \langle \hat{V} \rangle),$ where the factor $1/[(m_1+m_2)c^2\sqrt{\tau}]$ has been abbreviated as 
$\hat{ \mathcal E}$. Assume that the orthogonal components of the ``system" wave function, $|S_I\!\rangle $ and $|S_N\!\rangle$,  are well separated, and that the averaged effect of the potential, $V$, is approximately constant over the interacting component.  Since $V$ is assumed to be nonzero only in the region of interaction, the action of $ \hat{ \mathcal B}$ on the wave function during the entangling interaction can be calculated as follows: 
\begin{equation}\label{eqno8p4}
\begin{array}{ll} 
\hat{ \mathcal B} \; [(\alpha|S_I\!\rangle + \beta|S_N\!\rangle)\otimes|D\!\rangle]  \; = \;
\hat{ \mathcal E}(\alpha V |S_I\!\rangle|D_I\!\rangle -  \alpha^2 V (\alpha|S_I\!\rangle|D_I\!\rangle   
+ \beta|S_N\!\rangle |D_N\!\rangle)) \;\; \; \;   =  \\

\hat{ \mathcal E}\alpha V [(1-\alpha^2)|S_I\!\rangle|D_I\!\rangle -  \alpha\beta|S_N\!\rangle |D_N\!\rangle] \;\;   =   \; \; 

\hat{ \mathcal E}\alpha V [(\beta^2)|S_I\!\rangle|D_I\!\rangle -  \alpha\beta|S_N\!\rangle |D_N\!\rangle] \;\; \;    =   \\ 

\hat{ \mathcal E}\alpha\beta V [\beta|S_I\!\rangle|D_I\!\rangle -  \alpha|S_N\!\rangle |D_N\!\rangle].

\end{array}
\end{equation}

The resulting expression in the square brackets,  $\beta|S_I\!\rangle|D_I\!\rangle -  \alpha|S_N\!\rangle |D_N\!\rangle$, 
is clearly a normalized state that is orthogonal to the entangled state that results from the strictly linear evolution,  
$\alpha|S_I\!\rangle|D_I\!\rangle +  \beta|S_N\!\rangle |D_N\!\rangle$. This expression is multiplied by the coefficient, $\hat{ \mathcal E}\alpha\beta V$. The changes in  $\alpha$ and $\beta$ can be expressed as:  
$\epsilon_1 \; = \;  \hat{ \mathcal E} V \alpha\beta^2$  
and $\epsilon_2 \; = \;  \hat{ \mathcal E} V \alpha^2\beta.$\footnote{Since the term, 
	$ \;- \, \frac{1}{2}\hat{ \mathcal B}^\dagger \hat{ \mathcal B}  |\psi\,\rangle dt\;, $ is higher order in a small 	quantity, its effect can be neglected in this approximation.}

The deviation from linearity is calculated by taking the difference between the squared amplitudes of the cross terms\footnote{The differences in the cross terms are much easier to observe than the differences in the 	strictly symmetric and antisymmetric terms} in expressions, \ref{eqno8p3} (collapse), and \ref{eqno8p2} (no-collapse):  
\begin{equation}\label{eqno8p5}
\begin{array}{ll} 
(\alpha-\beta \pm (\epsilon_1+\epsilon_2))^2 - (\alpha-\beta )^2 \;  = \; 
(\epsilon_1+\epsilon_2)^2  \pm 2(\epsilon_1+\epsilon_2)(\alpha-\beta ) \;    = \;       \\

\hat{ \mathcal E}^2 V^2 ( \alpha^2\beta^4 + \alpha^4\beta^2 + 2\alpha^3\beta^3 )\; 
\pm \;2 \hat{ \mathcal E} V( \alpha\beta^2 + \alpha^2\beta)  (\alpha-\beta ).

\end{array}
\end{equation} 
This expression will be dominated by the second term, which is linear in the product, $\hat{ \mathcal E} V,$ (presumed to be quite small), except for extremely small values of $(\alpha-\beta )^2$ (the probability that is expected when there are no collapse effects). Since the positive and negative excursions in this term cancel, any collapse effects would tend to be masked. Therefore, the observability of possible deviations from linear evolution is maximized when $\alpha^2=\beta^2 = 1/2,$ and it depends on the magnitude of the term, 
$ \hat{ \mathcal E}^2 V^2 ( \alpha^2\beta^4 + \alpha^4\beta^2 + 2\alpha^3\beta^3).$ The expression in the parentheses involving $\alpha$ and $\beta$ sums to $1/2$, but the expression multiplying it, $ \hat{ \mathcal E}^2 V^2, $ is quadratic in a small quantity. The very small value of this term highlights what is probably the biggest obstacle to observing any deviations from strictly linear evolution. But there are 
other serious challenges as well.

The strong coupling that would be required to establish significant entanglement between the particles means that the particles would probably have to be charged. This would make it very difficult to maintain the precise control 
that would be required in order to observe any collapse effect. Substantial progress has been made in implementing double-slit quantum erasers using electronic Mach-Zehnder interferometers 
(EMZI)\cite{EMZI,EMZI_QE_1,EMZI_QE_2,EMZI_QE_3,EMZI_QE_4,EMZI_QE_5}, but further advances would be needed to carry 
out the sort of tests described here.

Double-slit quantum erasers using photons have been  implemented\cite{Scully_exp,Walborn}, and these allow for much more precise control. However, the separation of the wave functions into the branches between which the interference effects are observed does not correspond to the division into interacting and noninteracting components. Therefore no significant shift of amplitude between the interfering branches is expected based on the collapse equation that has been proposed. Although photons do not (typically) interact, it is conceivable that an experiment with a mediating system used to entangle two photons could eventually be designed.

It is possible that experimental techniques will advance enough so that tests could be carried out in the not-too-distant future. Another way in which to evaluate the proposed  hypothesis could be to show that it explained some otherwise unexplained naturally occurring phenomena. Because the collapse effects are assumed to scale with the strength of the interaction, phenomena involving the strong interaction might be good places to look.

\section{Discussion}
\label{sec:9}

The lack of an adequate explanation for wave function collapse leaves a huge gap in the logical structure of contemporary physics. The experimental foundation on which that structure rests would largely dissolve without the projection postulate and the Born probability rule. There is no way to relate particular experimental results to theoretical predictions without invoking these interpretive principles.

Bohr tried to avoid the logical problem by arguing that the predictions of quantum theory were invariant with respect to the point at which the collapse is assumed to take place. But Bell demonstrated very clearly that Bohr's claim was incorrect. By considering a simple change in the basis in which a system is measured he showed that there is, in principle, \textit{always} a discrepancy between some of the predictions of linear evolution and those of collapse. Given the current efforts to construct quantum computers, which address essentially the same kinds of issues as 
Bell's gedanken experiment, it is conceivable that tests of the general sort that he envisioned could eventually be carried out.

The main reason for the lack of an explanation for wave function collapse is that its nonlocal nature makes it very hard to incorporate it into a theoretical framework that adheres rigidly to a spacetime ontology based on classical relativistic notions. The refusal to challenge the ontological status of spacetime is unreasonable, given the readiness of so many to question the reality of other fundamental features of physics, and it runs counter to the methodology that Einstein employed in formulating the original theory. That methodology was based on the idea that the concepts used to describe space and time must be defined in terms of relationships among physical processes and entities.

To tame the nonlocality of wave function collapse Einstein's two original postulates were, implicitly, supplemented by a third. But this went largely unnoticed for a long time due to several factors. The Born  probability rule was originally proposed for reasons that were not clearly connected to any relativistic considerations. It was only realized later that it is necessary in order to avoid an open conflict between nonrelativistic quantum theory and the  prohibition of superluminal information transmission. The requirement of local commutativity, which plays an analogous role in quantum field theory, was not formulated until the Copenhagen interpretation had become so firmly fixed in the minds of most physicists that it was viewed as simply a ``translation" of Einstein's original postulate governing the speed of light into quantum language.

The eventual recognition that the supplemental rules are not equivalent to Einstein's postulate has led many to regard the no-superluminal-signaling principle as just a ``fall back" position that serves to maintain a ``peaceful coexistence" 
between quantum theory and relativity. The viewpoint advocated here is that this principle captures much of what is essential to both of these branches of contemporary physical theory, and that we should  try to explain it in terms of 
fundamental features of nature. It is notable that the assumption that wave function collapse is a genuine physical process opens the door to such an explanation.

Gisin has shown that wave function collapse implies that if superluminal signaling at the observable level is to be ruled out, then there must be nondeterministic effects at the elementary level. The lack of complete determinism places lower limits on the size of information processing systems. This points the way to an understanding of both information and the no-signaling principle in fundamental physical terms. 

We first stipulate that information is a reproducible and referential property of \textit{physical} systems, and recognize that acquisition of information that a random event (such as wave function collapse) has occurred in a spacelike-separated region would violate the no-superluminal signaling principle. To prevent such violation there must be some constraints on the manner in which wave function collapse is induced. It must be induced by the same physical processes by which  information is acquired, that is, correlating interactions. If the acquisition of information that a collapse has occurred is, itself, capable of inducing that collapse then no (new) information about spacelike-separated events can be obtained by the acquisition process.

A coherent account of nondeterministic, nonlocal effects can be most easily constructed by assuming a global sequencing relation among them. This sequencing remains unobservable because of the lower limits on the size of physical systems 
that can instantiate stable, reproducible information. The lack of observability leaves the relativistic transformation properties of the mathematical description of spacetime intact. But that mathematical structure is assigned a different ontological status within the overall theoretical framework.

The lack of complete determinism means that there are real physical processes that are not observable in detail. But it is still possible to characterize some of the general features of those processes. We can describe, in general terms, the stochastic evolution of quantum states with respect to a global sequencing parameter. The nonrelativistic stochastic collapse equations proposed by a number of authors can be adapted to this purpose by considering the special case in which the sequencing parameter coincides with the time in some (unknown) inertial frame. 

The connection between collapse and entangling interactions suggests that the collapse operator be defined in terms of the interaction potential energy operator, divided by the sum of masses of the interacting systems. This induces collapse in a natural way, and the effects scale with the entanglement generated by the interaction. This explanation parallels the conventional macroscopic account of projection induced by measurement, and like that account, at the observable level, it maintains conservation of energy, momentum, and related quantities\cite{Gillis_2}. Since it ties (genuine) wave function collapse to the processes typically associated with decoherence, many of the observable consequences of the proposal track those of decoherence accounts.

The hypothesis proposed here has both negative and positive implications regarding the prospects for experimental confirmation. On the negative side, the assumption that collapse effects are induced by the kinds of interactions usually associated with decoherence means that deviations from linearity would be quite difficult to observe. On the plus side, the direct connection between entangling interactions and collapse effects indicates that those effects can be brought under the control of experimenters. The stochastic equation based on the hypothesis also suggests 
experimental strategies that can maximize the chances of observing collapse effects. Definitive tests will require considerable work on both experimental and analytical fronts.

Attempts to incorporate an account of objective wave function collapse into the mathematical structure of contemporary theory have been one of the least favored approaches to dealing with the conceptual problems surrounding quantum measurement. This is true among both physicists generally and also those who focus on quantum foundations. The reluctance to consider this approach appears to stem from a deep ambivalence about the nondeterministic and nonlocal aspects of quantum phenomena. While there is widespread acknowledgment that these features characterize quantum theory in some fashion, the changes that would be required in both the mathematical framework and our understanding of physical theory have deterred most from exploring this path. What I have tried to show here is that 
treating wave function collapse as a real phenomenon can open the way to a clear understanding of the no-signaling principle in physical terms, and this new understanding imposes very useful constraints on the way in which collapse 
effects originate.


\begin{thebibliography}{99}
	
	
	
	\bibitem{Bell_CH} Bell, J.S.:  On wave packet reduction in the Coleman-Hepp 
	model.  Helv. Phys. Act. \textbf{48}, 93-98 (1975); doi.org/10.5169/seals-114661. doi.org/10.1142/9789812795854 0077. Reprinted in Speakable 
	and Unspeakable in Quantum Mechanics, Revised edition (2004) pp. 48/49; ISBN-13: 978-0521523387    	ISBN-10: 0521523389.
	
	\bibitem{Hepp} Hepp, K.: Quantum theory of measurement and macroscopic observables. 
	Helv. Phys. Act. \textbf{45}, 237-248 (1972); 	doi.org/10.5169/seals-114381.
	
	\bibitem{Bohr_EPR} Bohr, N.: Can quantum-mechanical description of physical 
	reality be considered complete? Phys. Rev. \textbf{48}, 696 (1935);doi.org/10.1103/PhysRev.48.696.      
	
	\bibitem{Heisenberg_a} Heisenberg, W.: Wandlungen in den Grundlagen der 
	Naturwissenschaft. S. Hirzel Verlag, Zurich (1949); doi.org/10.1002/ange.19620741014;  ISBN 10: 3777603635   ISBN-13: 978-3777603636. 
	
	\bibitem{Heisenberg_b} Bacciagaluppi, G., Crull, E.: Heisenberg 
	(and Schr\"{o}dinger, and Pauli) on Hidden Variables. In 
	Stud. Hist. Phil. Mod. Phys. \textbf{40}, 374-382, Elsevier, Amsterdam, Netherlands, (2009); doi.org/10.1016/j.shpsb.2009.08.004.     
	
	\bibitem{Everett} Everett, H.: Relative state formulation of quantum mechanics. 
	Rev. Mod. Phys. \textbf{29}, 454 (1957); doi.org/10.1103/RevModPhys.29.454.  
	
	\bibitem{Zurek_Decoh} Zurek,W.H.  D.: Pointer basis of quantum apparatus: Into what mixture does the wave 	packet collapse? Phys. Rev. \textbf{D 24}, 1516 (1981); doi.org/10.1103/PhysRevD.24.1516.                      

  	\bibitem{Hartle} Hartle, J.B.: The quasiclassical realms of this quantum universe. 
	Foundations of Physics \textbf{41}, 982-1006 (2011); doi.org/10.1007/s10701-010-9460-0.
	
	\bibitem{deBroglie} de Broglie, L.: Tentative d'interpretation causale 
	non-lineaire de la mechanique ondulatoire. Gautier-Villars, Paris, (1956); (OCLC number  602849199).                                       
	
	\bibitem{Bohm} Bohm, D.: A suggested interpretation of quantum theory in terms 
	of 'hidden' variables, I. Phys. Rev. \textbf{85}, 166-179 (1952); doi.org/10.1103/PhysRev.85.166.                             
	
	\bibitem{Bohm_Hiley} Bohm, D., Hiley, B. J.: The Undivided Universe: An Ontological Interpretation of Quantum Theory. Routledge, New York (1993); ISBN-13: 978-0415121859   ISBN-10: 041512185X. 
	
	\bibitem{GRW} Ghirardi, G.C., Rimini, A., Weber, T.: Unified dynamics for 
	microscopic and macroscopic systems. Phys. Rev. \textbf{D34},  470-491  (1986); doi.org/10.1103/PhysRevD.34.470.
	
	\bibitem{Ghirardi_Bassi} Bassi, A., Ghirardi, G.C.: Dynamical Reduction Models. 
	Phys. Rep.  \textbf{379}, 257-427 (2003); doi: 10.1016/S0370-1573(03)00103-0.
	
	\bibitem{Pearle_1} Pearle, P.: How stands collapse I. 
	J. Phys. A: Math. Theor., \textbf{40}, 3189-3204 (2007); doi.org/10.1088/1751-8113/40/12/S18.        

	\bibitem{Pearle_2} Pearle, P.: How stands collapse II. In:  Myrvold,W.C., Christian, J. (eds.) Quantum Reality, Relativistic Causality, and Closing the Epistemic Circle, \textbf{73} of The Western Ontario 
	Series in Philosophy of Science. pp. 257-292.  Springer, Heidelberg (2009); doi.10.1007/978-1-4020-9107-0, ISBN 978-1-4020-9106-3.   
	
	\bibitem{Bedingham} Bedingham, D.J.: Relativistic State Reduction Dynamics.
	Foundations of Physics \textbf{41}, 686-704 (2011); doi.org/10.1007/s10701-010-9510-7.  
	
	
	\bibitem{Gao_1} Gao, S.: The Meaning of the Wave Function:
	In search of the ontology of quantum mechanics. 
	Cambridge University Press, Cambridge (2017); ISBN-13: 978-1107124356   ISBN-10: 1107124352. 
     
	\bibitem{Scully_Druhl} Scully, M.O., Druhl, K.:  Quantum eraser: A proposed photon correlation experiment concerning observation and "delayed choice" 
	in quantum mechanics. Phys. Rev.  \textbf{A25}, 2208  (1982); doi.org/10.1103/PhysRevA.25.2208.     
	
	\bibitem{Scully_ES_a} Scully, M.O., Englert, B.G., Schwinger, J.: Spin 
	coherence and Humpty-Dumpty. III. The effects of observation. 
	Phys.Rev. \textbf{A40}, 1775 (1989); doi.org/10.1103/PhysRevA.40.1775.
	
	\bibitem{Scully_ES_b} Scully, M.O., Englert, B.G., Walther, H.:  
	Quantum Optical Tests of Complementarity. Nature \textbf{351}, 111-116, London  (1991); doi.org/10.1038/351111a0.                      
	
	\bibitem{Scully_exp} Kim, Y-H, Rong Y., Kulik, S.P., Shih, Y.H., Scully, M.O.: 
	A Delayed Choice Quantum Eraser. Phys. Rev. Lett. \textbf{84}, 1-5 (2000);  doi:10.1103/PhysRevLett.84.1.     
	
	\bibitem{Walborn} Walborn, S., Terra Cunha, M.O., Padua, S., Monken, C.H.: 
	Double-slit quantum eraser. Phys. Rev. \textbf{A65}, 033818 (2002); doi: 10.1103/PhysRevA.65.033818.  
	
	\bibitem{Gisin_c} Gisin, N.: Stochastic quantum dynamics and relativity.  Helv. Phys. Act. \textbf{62}, 363-371 (1989); doi.org/10.5169/seals-116034.
	
	\bibitem{No_Clone} Wooters, W.K., Zurek, W.H.: A single quantum cannot be cloned. Nature \textbf{299}, 802-803 (1982); doi.org/10.1038/299802a0.                          
	
	\bibitem{Orig_Rel} Einstein, A.: On the Electrodynamics of Moving Bodies. Annalen der Physik \textbf{17}, 891-921 (1905);  doi:10.1002/andp.19053221004. 
	
	\bibitem{Pearle_1976} Pearle, P.: Reduction of the state vector by a nonlinear Schr\"{o}dinger equation. Phys. Rev. \textbf{D 13}, 857 (1976); doi.org/10.1103/PhysRevD.13.857. 
	
	\bibitem{Pearle_1979} Pearle, P.: Toward explaining why events occur. 
	Int. J. Theor. Phys. \textbf{18}, 489-518 (1979); doi.org/10.1007/BF00670504. 
		
	\bibitem{Gisin_1984} Gisin, N.: Quantum Measurements and Stochastic Processes.
	Phys. Rev. Lett. \textbf{52}, 1657 (1984); doi.org/10.1103/PhysRevLett.52.1657.
	
	\bibitem{Diosi_1} D\'{i}osi, L.: Continuous quantum measurement and the It\^{o} formalism. 	Phys. Lett. \textbf{129A}, 419-423 (1988); doi.org/10.1016/0375-9601(88)90309-X.	 
	
	\bibitem{Diosi_2} D\'{i}osi, L.: Quantum stochastic processes as models for state vector reduction. J. Phys. \textbf{A 21}, 2885-2898 (1988); doi:10.1088/0305-4470/21/13/013. 
	
	\bibitem{Diosi_3} D\'{i}osi, L.: Localized solution of a simple nonlinear quantum Langevin equation. Phys. Lett. \textbf{132A}, 233-236 (1988); doi.org/10.1016/0375-9601(88)90555-5.
	
	\bibitem{GPR} Ghirardi, G.C., Pearle, P., Rimini, A.: Markov-processes in Hilbert-space and continuous spontaneous localization of systems of identical particles. Phys. Rev. \textbf{A42},  78  (1990); doi.org/10.1103/PhysRevA.42.78. 
	
	\bibitem{Adler_Brun} Adler,S.L., Brun,T.A.: Generalized stochastic Schr\"{o}dinger equations for state vector collapse. J. Phys. \textbf{A 34}, 4797-4810 (2001);  doi: 10.1088/0305-4470/34/23/302. 
		
	\bibitem{Gillis_2} Gillis, E.J.: Interaction-Induced Wave Function Collapse Respects Conservation Laws. arXiv:1803.02687 v5  [quant-ph] (2018).   
		
	\bibitem{Bio_OReilly} O'Reilly, E.J., Olaya-Castro, A.:  
	Non-classicality of the molecular vibrations assisting exciton energy transfer at room temperature. 	Nature Communications \textbf{5}, 3012, (2014);  doi.org/10.1038/ncomms4012.          
	
	\bibitem{Gisin_a} Bruno, N., Martin, A., Sekatski,P., Sangouard,N., Thew, R., Gisin, N.: 	Displacement of entanglement back and forth between the micro and macro domain. Nature Physics \textbf{9}, 2681 (2013); doi:10.1038/nphys2681
	
	\bibitem{Vandersypen} Vandersypen, L.M.K., Steffen, M., Breyta, G., 
	Yannoni, C.S., Cleve, R., Chuang, I.L.: Implementation of a three-quantum-bit 
	search algorithm. Appl. Phys. Lett.\textbf{76}, 646 (2000); doi.org/10.1063/1.125846
	
	\bibitem{Steffen} Steffen, M., van Dam, W., Hogg, T., Breyta, G., Chuang, I.: 
	Experimental implementation of an adiabatic quantum optimization       
	algorithm. Phys. Rev. Lett. \textbf{90} (2003); doi.org/10.1103/PhysRevLett.90.067903.              
	
	\bibitem{Mermin} Mermin, N.D.: Quantum Computer Science: An Introduction, 
	Cambridge University Press, (2007);  doi.org/10.1017/CBO9780511813870. 
	
	\bibitem{Gillis_E} Gillis, E.J.: Causality, Measurement, and Elementary Interactions.  Foundations of Physics \textbf{41}, 1757-1785 (2011);  doi.org/10.1007/s10701-011-9576-x.  
	
	\bibitem{Svetlichny_a} Svetlichny, G.: Long Range Correlations and Relativity: 
	Metatheoretic Considerations. In:  Chybakolo, A.E., Smirnov-Rueda, R. (eds.)  
	Instantaneous Action-at-a-Distance in Modern Physics: "Pro" and "Contra", Pat II,Ch.10  Nova Science, Commack, N.Y., (1999);  ISBN 1560726989.   
	
	\bibitem{Svetlichny_b} Svetlichny, G.: Causality implies formal state collapse.     
	Foundations of Physics \textbf{33}, 641-655 (2003);  doi.org/10.1023/A:1023774721109.   

	\bibitem{Gleason} Gleason, A.M.: Measures on the closed subspaces of a Hilbert space.  Journal of Mathematics and Mechanics \textbf{6},  885-893 (1957); doi: 10.1512/iumj.1957.6.56050.     
	
	\bibitem{Born} Born, M.: On the quantum mechanics of collisions. Zeitschrift fur 
	Physik \textbf{37}, 863-67 (1926); doi:10.1007/BF01397477. (ISBN 0-691-08316-9.)
	
	\bibitem{Elitzur_1992} Elitzur, A. C.: Locality and indeterminism preserve 
	the second law. Phys. Lett. A  \textbf{167}, 335-340  (1992); doi.org/10.1016/0375-9601(92)90268-Q.  
	
	\bibitem{Popescu} Popescu, S., Rohrlich, D.: Quantum Nonlocality as an Axiom. 
	Foundations of Physics \textbf{24}, 379-385 (1994); doi.org/10.1007/BF02058098.                


	\bibitem{Elitzur_Dolev_1} Elitzur, A.C., Dolev, S.:  Quantum Phenomena within a New 
	Theory of Time.  In Elitzur A.C., Dolev S., Kolenda N. (eds.) Quo Vadis Quantum Mechanics?  Springer, Berlin Heidelberg (2005);  
	doi.org/10.1007/3-540-26669-0 17;ISBN 978-3-540-26669-3
		
	\bibitem{Qi-Ren} Zhang, Q.-R.:  Statistical Separability And The Consistency 
	Between Quantum Theory, Relativity, And The Causality. Communications in Theoretical Physics \textbf{47}(5):826-828 (2007); doi.org/10.1088/0253-6102/47/5/012.
	
	\bibitem{Masanes} Masanes, Ll., Acin, A., Gisin, N.: General Properties of 
	Nonsignaling Theories. Phys. Rev. \textbf{A73}, 012112 (2006); doi.org/10.1103/PhysRevA.73.012112.
	
	\bibitem{Bell_1} Bell, J.S.: On the Einstein Podolsky Rosen paradox. Physics 
	\textbf{1}, 195-200 (1964);   doi: 10.1103/PhysicsPhysiqueFizika.1.195.  Reprinted in Speakable and Unspeakable in Quantum Mechanics, Revised edition (2004) pp. 14-21; ISBN-13: 978-0521523387    	ISBN-10: 0521523389.
	
	\bibitem{EPR} Einstein, A., Podolsky, B., Rosen, N.: Can quantum-mechanical 
	description of reality be considered complete?. Phys. Rev. \textbf{47} 777 (1935); doi.org/10.1103/PhysRev.47.777. 
	
	\bibitem{Bell_LNC} Bell, J.S.: La Nouvelle Cuisine. In: Sarlemijn,  A.,  
	Kroes, P. (eds.) Between Science and Technology, p.000. Elsevier Science 
	Publishers, Oxford (1990); eBook ISBN: 9780444597304. Reprinted in Speakable 
	and Unspeakable in Quantum Mechanics, Revised edition (2004) pp. 232-248; ISBN-13: 978-0521523387    	ISBN-10: 0521523389.                      
	
	\bibitem{Landauer} Landauer, R.: Irreversibility and heat generation in the computing process. 	IBM Journal of Research and Development, \textbf{5}, 183-191 (1961);  doi: 10.1147/rd.53.0183.
	
	\bibitem{Bennett_1973} Bennett, C.H.: Logical reversibility of computation. IBM Journal of Research and Development, \textbf{17}, 525-532 (1973);  doi: 10.1147/rd.176.0525.
	
	\bibitem{Bennett_1982} Bennett, C.H.: The thermodynamics of computation - a review. International Journal of Theoretical Physics \textbf{21}, 905-940 (1982);  doi.org/10.1007/BF02084158.
	
	\bibitem{Bennett_Landauer} Bennett, C.H., Landauer, R.: The Fundamental Physical Limits of Computation. 	Scientific American, \textbf{263},3,48 (1985); DOI: 10.1038/scientificamerican0785-48.
	
	\bibitem{Landauer_size} Landauer, R.: Computation: A fundamental physical view. 
	Phys. Scr., \textbf{35}, (88-95) (1987); doi.org/10.1088/0031-8949/35/1/021.                    
	
	\bibitem{Landauer_99} Landauer, R.:  Information is a physical entity. Physica A, \textbf{263}, 63-67 (1999); doi.org/10.1016/S0378-4371(98)00513-5.
	
	\bibitem{Bennett_2003} Bennett, C.H.: Notes on Landauer's principle, reversible computation, and Maxwell's Demon. Studies Hist. Philos. Modern Phys. \textbf{34} 501-510  (2003); doi.org/10.1016/S1355-2198(03)00039-X.
	
	\bibitem{Wheeler_it_bit} Wheeler, J. A.: Information, physics, quantum: the search for links. 	\textit{Proceedings III International Symposium on Foundations of Quantum Mechanics, Tokyo}, 354-368  (1989);    
	ISBN-10: 4890270035    ISBN-13: 9784890270033.
	
	\bibitem{Maudlin} Maudlin, T.: Quantum Non-Locality and Relativity. 2nd Edition, 
	Blackwell Publishers Inc., Malden (2002);  ISBN-13: 978-0631232209   ISBN-10: 0631232206.
		
	\bibitem{Bohm_Rel}D\"{u}rr, D., Goldstein, S., Norsen, T., Struyve, W., Zanghi, N.: 
	Can Bohmian mechanics be made relativistic?   Proc Math Phys Eng Sci. \textbf{470} 2162 (2014); doi:  10.1098/rspa.2013.0699    
	
	\bibitem{Struyve_Tumulka} Struyve, W., Tumulka, R.: 
	Bohmian Mechanics for a Degenerate Time Foliation.  Quantum Studies: Mathematics and Foundations  \textbf{2} 349-358 (2015); doi.org/10.1007/s40509-015-0048-4.      
	
	\bibitem{Bancal_Gisin} Bancal, J-D., Pironio,S. , Acin, A., Liang,  Y-C.,  Scarani, V., Gisin, N.: Quantum non-locality based on finite-speed causal influences leads 
	to superluminal signaling. Nature Physics \textbf{8}, 867-870 (2012);  doi.org/10.1038/nphys2460.
	
	\bibitem{Gisin_f} Gisin, N.: Quantum correlations in Newtonian space and time: 
	Arbitrarily fast communication or nonlocality.  In: Struppa, D.C. and Tollaksen, J.M, (eds.) Quantum Theory: A Two-Time Success Story, Yakir Aharonov Festschrift. 185-204. Springer, (2013);  ISBN 978-88-470-5217-8.   
	
	\bibitem{Bell_LI} Bell, J.S.:  Towards an exact quantum mechanics. In: Deser,S., Finkelstein,R.J., (eds.)  Themes in contemporary physics, II: essays in honor of Julian Schwinger's 70th birthday, World scientific, Singapore (1989); ISBN-10: 997150961X  	ISBN-13: 978-9971509613.
		
	\bibitem{Tumulka} Tumulka, R.: A Relativistic Version of the Ghirardi$-$Rimini$-$Weber Model. J. Stat. Phys. \textbf{125},  821-840 (2006); doi.org/10.1007/s10955-006-9227-3. 
	
	\bibitem{Pearle_RDCM} Pearle, P.: Relativistic dynamical collapse model.
	Phys. Rev. D \textbf{91}, 105012 (2015);  doi: 10.1103/PhysRevD.91.105012.	     
		
	\bibitem{von_Neumann} von Neumann, J.: Mathematische Grundlagen der 
	Quantenmechanik.  Springer, Berlin (1932); ISBN 978-3-540-59207-5.	
	
	
	\bibitem{Bassi_Ghi} Bassi, A., Ghirardi, G.C.: Dynamical Reduction Models.
	Phys. Rep. \textbf{379} 257-426 (2003);  doi:10.1016/S0370-1573(03)00103-0. 	


	\bibitem{Gao_2} Gao, S.: A discrete model of energy-conserved wavefunction collapse.
	Proceedings of the Royal Society \textbf{A 469}, 20120526. (2013);  

	
	\bibitem{EMZI} Ji, Y., Chung, Y., Sprinzak, D. , Heiblum, M., Mahalu, D. , Shtrikman, H.:  An Electronic Mach-Zehnder Interferometer. Nature \textbf{422}, 415 (2003); doi:10.1038/nature01503.
	
	  
	\bibitem{EMZI_QE_1} Kang, K. : Electronic Mach-Zehnder quantum eraser. Phys. Rev. \textbf{B 75}, 125326 (2007); doi:10.1103/PhysRevB.75.125326.  

	\bibitem{EMZI_QE_2} Kang, K., Lee, K.H.: Violation of Bell's inequality in electronic Mach-Zehnder interferometers. Physica E \textbf{40}, 1395-1397 (2008);  doi:10.1016/j.physe.2007.09.124. 
	
	\bibitem{EMZI_QE_3} Dressel,J., Choi,Y., Jordan,A. N.: Measuring which-path information with coupled electronic Mach-Zehnder interferometers. Phys. Rev. \textbf{B 85}, 045320 (2012);  doi:10.1103/PhysRevB.85.045320.

	
	\bibitem{EMZI_QE_4} Buscemi, F., Bordone, P., Bertoni, A.: Electron interference and entanglement in coupled 1D systems with noise. Eur. Phys. J. \textbf{D 66} 312  (2012); doi:10.1140/epjd/e2012-30469-5.
	
	\bibitem{EMZI_QE_5} Weisz, E., Choi, H. K., Sivan, I., Heiblum, M., Gefen, Y., Mahalu, D., Umansky, V.: An Electronic Quantum Eraser. Science \textbf{344} (2014);  doi: 10.1126/science.1248459.        
	
	
	
\end{thebibliography}
\end{document}